%Paper: gr-qc/9404065
%From: carroll@marie.mit.edu (Sean Carroll)
%Date: Sun, 1 May 1994 16:35:00 -0400

%%%%%%%%%%%%%%%%%%%%%%%%%%%%%%%%%%%%%%%%%%%%%%%%%%%%%%%%%%%%%%
%
%   This paper comes with a uuencoded compressed file with
%   directions for unpacking.  It creates nine figure files
%   which will be automatically included when the paper is
%   printed (epsf etc. not required).  There are also a set
%   of switches below, which allow you to choose various
%   printing styles.
%
%%%%%%%%%%%%%%%%%%%%%%%%%%%%%%%%%%%%%%%%%%%%%%%%%%%%%%%%%%%%%%

% ``ENERGY-MOMENTUM RESTRICTIONS ON THE CREATION OF GOTT TIME
%    MACHINES''
%  by Sean M. Carroll, Edward Farhi, Alan H. Guth and Ken D. Olum
%
%  April 29, 1994: Final version

% Switches for printing styles:
   \newif\ifsansserif \newif\iftwoup \newif\ifdoublespace
   \newif\ifam \newif\ifaddmargin \newcount\figurestyle
\sansseriftrue     % Selects sans serif vs roman
\twouptrue         % Selects two-up style
\doublespacefalse  % Selects double spacing vs single spacing
\amfalse           % Selects "am" fonts vs "cm" fonts
\addmarginfalse    % If true, adds a 1 inch margin (for Imagen printers)
\figurestyle=3     % Selects treatment of figures:
                   %   1: no figures, captions at end
                   %   2: captions in text, no figures
                   %   3: captions and figures embedded in text

\def\draftdate{}

\newdimen\fullhsize \newbox\leftcolumn
\iftwoup
   \magnification=1000
   
   \fullhsize=10truein
   \def\fulline{\hbox to \fullhsize}
   \hfuzz=5pt
   \hsize=4.75truein \hoffset=-0.54truein
   \vsize=6.5truein \voffset=0truein
   \let\lr=L
   \output={\if L\lr
      \global\setbox\leftcolumn=\columnbox \global\let\lr=R
      \else\doubleformat \global\let\lr=L\fi
      \advancepageno
      \ifnum\outputpenalty>-20000 \else\dosupereject\fi}
   \def\doubleformat{\shipout\vbox{\makeheadline
      \fulline{\box\leftcolumn\hfil\columnbox} }}
   \def\columnbox{\leftline{\vbox{{\pagebody\makefootline}}}}
\else
   \magnification 1200
   \hsize=6.4 truein \vsize=8.9 truein
   \hoffset=0 truein \voffset=0 truein
\fi

\ifaddmargin
   \advance\hoffset by 1truein \advance\voffset by 1 truein
\fi

% Font definitions:
\ifsansserif
   \ifam
      \font\normal=amss10
      \font\bf=amssbx10
      \font\it=amssi10
      
      \iftwoup
         \font\small=amss10 scaled 800
      \else
         \font\small=amss10 scaled 833
      \fi
      \font\bigbf=amssbx10 scaled 1200
      \font\bigbigbf=cmssbx10 scaled 1440
      % Define math fonts that are a little bolder than normal:
         \font\tenss=amr8 scaled 1200
         \font\sevenss=amr6 scaled 1200
         \font\fivess=amr5
         \font\tenssi=ammi8 scaled 1200
         \font\sevenssi=ammi6 scaled 1200
         \font\fivessi=ammi5
         \font\tensssy=amsy8 scaled 1200
         \font\sevensssy=amsy6 scaled 1200
         \font\fivesssy=amsy5
         \textfont0=\tenss \scriptfont0=\sevenss
         \scriptscriptfont0=\fivess
         \textfont1=\tenssi \scriptfont1=\sevenssi
         \scriptscriptfont1=\fivessi \textfont2=\tensssy
         \scriptfont2=\sevensssy \scriptscriptfont2=\fivesssy
   \else
      \font\normal=cmss10
      \font\bf=cmssbx10
      \font\it=cmssi10
      
      \font\small=cmss8
      \font\bigbf=cmssbx10 scaled 1200
      \font\bigbigbf=cmssbx10 scaled 1440
   \fi
\else
   \ifam
      \font\normal=amr10
      \font\bf=ambx10
      \font\it=amti10
      
      \font\small=amr10 at 10 truept
      \font\bigbf=ambx10 scaled 1200
      \font\bigbigbf=ambx10 scaled 1440
   \else
      \font\normal=cmr10
      \font\bf=cmbx10
      \font\it=cmti10
      
      \font\small=cmr10 at 10 truept
      \font\bigbf=cmbx10 scaled 1200
      \font\bigbigbf=cmbx10 scaled 1440
   \fi
\fi
\normal

%Modification of format for footnotes:
\catcode`@=11
 \def\vfootnote#1{\insert\footins\bgroup
    \interlinepenalty\interfootnotelinepenalty
    \splittopskip\ht\strutbox % top baseline for broken footnotes
    \splitmaxdepth\dp\strutbox \floatingpenalty\@MM
    \leftskip\z@skip \rightskip\z@skip \spaceskip\z@skip
    \xspaceskip\z@skip \footformat \footfont
    \ifdim \parindent < 10pt \parindent=10pt \fi
    \textindent{#1}\footstrut\futurelet\next\fo@t}
 \def\footformat{} \def\footfont{}
 %To change the linespacing in footnotes:
     \ifdoublespace
        \def\footformat{\baselineskip=24pt}
        \setbox\strutbox=\hbox{\vrule height20.5pt depth3.5pt width0pt}
     \else
        \def\footformat{\baselineskip=12pt}
     \fi
 %To change the space between text and footnotes on a page:
        \skip\footins=18pt
 %To make the footnote rule the same width as the page:
        \def\footnoterule{\kern-3pt
        \hrule width \the\hsize \kern 2.6pt}
\catcode`@=12 % at signs are no longer letters

\def\pagenumbers{\footline={\hss\normal\ifnum\pageno>0
   \folio\fi\hss}}
\pagenumbers \pageno=0

\def\.{.{\spacefactor=3000}}  % Period with long space; use with
%   abbreviations at end of sentences.  Example: MIT\.
\def\shortperiod{\sfcode`.=1000} %Gives short spaces following periods.

\def\head#1{\par \ifnum \lastpenalty < 10000 \goodbreak \fi
   \bigskip {\baselineskip=15pt \hangindent=30pt {\noindent
   {\bigbf #1} \par \hangindent=0pt}\nobreak \medskip \nobreak}}
   % Big Boldface headlines at left margin
\def\subhead#1{\par \ifnum \lastpenalty < 10000 \goodbreak \fi
   \medskip {\baselineskip=12pt \hangindent=30pt {\noindent
   {\bf #1} \par \hangindent=0pt}\nobreak \smallskip \nobreak}}
   % Boldface headlines at left margin

% The following macros are equivalent to adding
%   \noalign{\smallskip} between any two lines of \eqalign,
%   \eqalignno, \leqalignno, \displaylines, and the \eqcenter
%   macro defined below.  \matrix and \pmatrix are not affected.
\jot = 6pt   % increased from default of 3pt
\catcode`@=11
\def\cases#1{\left\{\,\vcenter{\normalbaselines\openup3pt\m@th
    \ialign{$##\hfil$&\quad##\hfil\crcr#1\crcr}}\right.}
    % \openup3pt was added
\catcode`@=12

% The following macros redefine \eqno and \eqalignno, so that the
%    output is printed in \normal font, (either \rm or
%    \sansserif).  This only makes a difference when the equation
%    numbers contain latin letters, such as 1a.
\let\oldeqno = \eqno
\def\eqno#1$${\oldeqno{\hbox{\normal #1}}$$}
\catcode`@=11
\def\eqalignno#1{\displ@y \tabskip\centering
  \halign to\displaywidth{\hfil$\@lign\displaystyle{##}$\tabskip\z@skip
    &$\@lign\displaystyle{{}##}$\hfil\tabskip\centering
    &\llap{$\@lign{\hbox{\normal ##}}$}\tabskip\z@skip\crcr
    #1\crcr}}
\catcode`@=12 % at signs are no longer letters

\def\standardbaselineskip{\ifdoublespace \baselineskip=24pt plus
   1pt minus 1pt \else \baselineskip=12pt \fi}
\standardbaselineskip
\def\standardparskip{\ifdoublespace \parskip=0pt plus 1 pt
   \else \parskip=4pt plus 2pt minus 2pt\fi}
\standardparskip

%The following footnote and reference macros implement what is
%   the new policy for Phys. Rev.  According to a recent
%   announcement,
%      ``References and footnotes are separately available.
%   References are designated by on-line numbers in square
%   brackets.  Footnotes are placed at the bottom of the page on
%   which they are cited and are designated by superscript
%   numbers and numbered consecutively throughout the paper.''
%      I am using the macro \rf{} for references, and footnotes
%   will be entered by \fn{<N>}{<text>}, where N is the number.
%   The superscript mechanism arranges for the extra
%   end-of-sentence space to be included in the space following
%   the footnote reference.  It also arranges for a dash to be
%   printed as if it were a double dash surrounded by thinspaces.
\def\rf#1{\unskip \ [#1]} % Puts a space before [ ] if none exists
\def\ddash{--}
{\catcode`-=\active
   \gdef\sup{\bgroup \catcode`-=\active \fnactive}
   \gdef\fnactive#1{\let\asf=\empty
      \ifhmode\xdef\asf{\spacefactor=\the\spacefactor}\fi
      {\def-{$\,$\ddash$\,$}%
      \baselineskip=10pt $^{\hbox{\small #1}}$}\egroup\asf{}}
}
\def\fn#1#2{\footnote{\sup{#1}}{#2}}

\ifnum\figurestyle>1    % Captions in text for preprint mode only.
  \def\figcapt#1#2{{\baselineskip=10pt \narrower\smallskip\noindent
     Fig.\ #1\ #2\smallskip}}
  \def\figureinserta#1#2#3#4#5#6{\topinsert
     \ifnum\figurestyle=3
        \vbox to #4{} \nointerlineskip \centerline{\hbox to
        #2{\includegraphics{#6}\hfill}}
     \fi
     \vskip #3 \figcapt{{#1}:}{#5}\endinsert}
  \def\figureinsert#1#2#3#4#5#6{\ifhmode\vadjust{\figureinserta{#1}
     {#2}{#3}{#4}{#5}{#6}}\ignorespaces \else
     \figureinserta{#1}{#2}{#3}{#4}{#5}{#6} \fi}
  \def\figurepageinserta#1#2#3#4#5#6{\pageinsert \vbox{} \vfill
     \ifnum\figurestyle=3
        \vbox to #4{} \nointerlineskip \centerline{\hbox to
        #2{\includegraphics{#6}\hfill}}
     \fi
     \vskip #3 \figcapt{{#1}:}{#5}\vfill\endinsert}
  \def\figurepageinsert#1#2#3#4#5#6{\ifhmode\vadjust{%
     \figurepageinserta{#1}{#2}{#3}{#4}{#5}{#6}}\ignorespaces \else
     \figurepageinserta{#1}{#2}{#3}{#4}{#5}{#6} \fi}
  \def\captioninserta#1#2#3{\topinsert \vskip #2 \figcapt{{#1}:}{#3}
     \endinsert}
  \def\captioninsert#1#2#3{\ifhmode\vadjust{\captioninserta{#1}
     {#2}{#3}}\ignorespaces \else \captioninserta{#1}{#2}{#3} \fi}
\else
  \def\figureinsert#1#2#3#4#5#6{\ignorespaces}
  \def\figurepageinsert#1#2#3#4#5#6{\ignorespaces}
  \def\captioninsert#1#2#3{\ignorespaces}
\fi

% THE FOLLOWING MACROS ALLOW BOTTOMINSERT INSERTIONS, IN THE FORMAT
% \bottominsert <text> \endbottominsert
\newinsert\botins
\skip\botins=\bigskipamount
\count\botins=1000       % magnification factor (1 to 1)
\dimen\botins=\maxdimen  % no limit per page
\catcode`@=11
\def\bottominsert{\@ins}
\def\endbottominsert{\egroup % finish the \vbox
   \insert\botins{\penalty100 % floating insertion
   \splittopskip\z@skip
   \splitmaxdepth\maxdimen \floatingpenalty\z@
   \box\z@}\endgroup}
\def\pagecontents{\ifvoid\topins\else\unvbox\topins\fi
  \dimen@=\dp\@cclv \unvbox\@cclv % open up \box255
  \ifvoid\botins\else
    \vskip\skip\botins
    \unvbox\botins \fi
  \ifvoid\footins\else % footnote info is present
    \vskip\skip\footins
    \footnoterule
    \unvbox\footins\fi
  \ifr@ggedbottom \kern-\dimen@ \vfil \fi}
\catcode`@=12
% END OF BOTTOMINSERT MACROS

\def\references{\parskip=\medskipamount
   \shortperiod \def\jf{\it}}  % jf = font for journal names
\def\ref#1{\par \hangafter 1 \hangindent 35 pt \noindent
    \hbox to 20pt{\hfil #1.}\ \ignorespaces}
\def\captions{\parskip=\medskipamount}
\def\capo#1#2{\par\hangafter 1 \hangindent 20 pt {\par
    \noindent Figure {#1}:  {#2}}}

\def\gmatrix#1{\left\lgroup \matrix{#1} \right\rgroup}
\def\sec{Sec.~}

\def\fig{Fig.~}

\def\eq{Eq.~}
\def\eqs{Eqs.~}
\def\eqpar#1{(#1)} % Used to put a number in parentheses and to
                   % indicate to the renumbering program that it
                   % is an equation number.
\def\Ref#1{Ref.~[#1]}
\def\refbrack#1{[#1]} % Used to put a number in brackets to
                      % indicate to the renumbering program that
                      % it is a reference number.

\def\svec#1{\skew{-2}\vec#1}
\def\vecxi{\svec \xi}

\font\scr=cmbsy10
\def\J{\hbox{\scr J}}
\def\Tr{\mathop{\rm Tr}\nolimits}
\def\p0{(n^0)^2-1}

\def\ucg{\widetilde{\hbox{SU}}\hbox{(1,1)}}
\def\diag{\,{\rm diag}\,}
\def\gmatrix#1{\left\lgroup \matrix{#1} \right\rgroup}
             % Boldfaced italics, for emphasis
\def\IR{\relax{\rm I\kern-.18em R}}   % Blackboard font R
% \bar V, but with wider bar:
\def\vbar{\hbox{\vbox{\nointerlineskip\moveright 1pt
   \hbox{\leaders\hrule\hskip 6pt}\vskip -4pt \hbox{$V$}}}}

\def\nulltest{}
\vbox{} \vskip -\topskip \vskip -\baselineskip
\vbox to 0pt{\vskip -18pt \ifx \draftdate \nulltest \else
   \line{\hfill DRAFT: \draftdate}\fi \vfill}
\baselineskip 12pt plus 1pt minus 1pt
\vskip 0pt plus 2fill
 at 17.28truept
\iftwoup
   \centerline{\bigbf ENERGY-MOMENTUM RESTRICTIONS ON}
   \medskip
   \centerline{\bigbf THE CREATION OF GOTT TIME MACHINES}
\else
{\baselineskip=20pt
   \centerline{\bigbigbf ENERGY-MOMENTUM RESTRICTIONS ON}
   \medskip
   \centerline{\bigbigbf THE CREATION OF GOTT TIME MACHINES}
}
\fi
\vskip 0pt plus 2fill
\centerline{Sean M. Carroll,\footnote{$^{*}$}{This work was
supported in part by funds provided by the U.S. National
Aeronautics and Space Administration (NASA) under contracts
NAGW-931 and NGT-50850, and the U.S. National Science Foundation
under grant PHY/9206867.}$^{,\hbox{\S}}$ Edward
Farhi,$^{\hbox{\S}}$ Alan H. Guth,\footnote{$^{\hbox{\S}}$}{This
work was supported in part by funds provided by the U.S.
Department of Energy (D.O.E.) under contract
\#DE-AC02-76ER03069.} and Ken D. Olum\footnote{$^{\dag}$} {This
work was supported in part by funds provided by the Texas
National Research Laboratory Commission under grant
\#RGFY93-278C.}}

\bigskip
\centerline{{\it Center for Theoretical Physics}}
\centerline{\it Laboratory for Nuclear Science}
\centerline{\it and Department of Physics}
\centerline{\it Massachusetts Institute of Technology}
\centerline{\it Cambridge, Massachusetts\ \ 02139\ \ \ U.S.A.}
\vskip 0pt plus 5fill
\centerline{Submitted to {\it Physical Review D}}
\vskip 0pt plus 2fill
\centerline{}
\noindent CTP\#2252, gr-qc/9404065 \hfill April 1994
\smallskip
\noindent{This paper replaces CTP\#2117, ``Gott Time Machines
Cannot Exist in an Open (2+1)-Dimensional Universe with Timelike
Total Momentum,'' by S. M. Carroll, E. Farhi, and A. H. Guth.}
This version is precise about the meaning of ``open universe,''
and proves that if the total momentum of such a universe is
timelike then the total deficit angle is no more than $2 \pi$.
\eject

\standardbaselineskip
{\baselineskip 14pt
\centerline{\bf ENERGY-MOMENTUM RESTRICTIONS ON}
\centerline{\bf THE CREATION OF GOTT TIME MACHINES}
}
\bigskip
\centerline{Sean M. Carroll, Edward Farhi, Alan H. Guth,
and Ken D. Olum}
\bigskip\bigskip

\centerline{\bigbf ABSTRACT}
\medskip

The discovery by Gott of a remarkably simple spacetime with
closed timelike curves (CTC's) provides a tool for investigating
how the creation of time machines is prevented in classical
general relativity.  The Gott spacetime contains two infinitely
long, parallel cosmic strings, which can equivalently be viewed
as point masses in (2+1)-dimensional gravity.  We examine the
possibility of building such a time machine in an open universe.
Specifically, we consider initial data specified on an edgeless,
noncompact, spacelike hypersurface, for which the total momentum
is timelike (i.e., not the momentum of a Gott spacetime).  In
contrast to the case of a closed universe (in which Gott pairs,
although not CTC's, can be produced from the decay of stationary
particles), we find that there is never enough energy for a
Gott-like time machine to evolve from the specified data; it is
impossible to accelerate two particles to sufficiently high
velocity.  Thus, the no-CTC theorems of Tipler and Hawking are
enforced in an open (2+1)-dimensional universe by a mechanism
different from that which operates in a closed universe. In
proving our result, we develop a simple method to understand the
inequalities that restrict the result of combining momenta in
(2+1)-dimensional gravity.

\bigskip
\head{I. INTRODUCTION}

Absent some restriction on boundary conditions and energy
sources, it is possible for the spacetime metric of general
relativity to wreak havoc with our intuitive notions of ``going
forward in time.'' We can imagine metrics in which the worldline
of a test particle, locally restricted to the interior of its
forward light cone, can loop around to intersect itself --- a
closed timelike curve (CTC)\.  Indeed, it is easy to construct
solutions to Einstein's equations that exhibit such behavior
\rf{1-3}.

Nonetheless, due to the causal paradoxes associated with such a
time machine, it is tempting to believe that CTC's exist only in
spacetimes that are in some way pathological.  That is, we would
expect that the laws of physics somehow act to prevent the
occurrence of CTC's in the real universe.  This expectation has
been dubbed the ``Chronology Protection Conjecture'' \rf{4}.

In the context of classical general relativity, a counterexample
to the chronology protection conjecture would be a solution to
Einstein's equations that describes the creation of CTC's, using
only ordinary materials, in a local region of a spacetime that is
free of CTC's in the past.  There is evidence, in the form of
theorems proven by Tipler \rf{5} and Hawking \rf{4}, that no such
solutions exist.  These results demonstrate that CTC creation in
a local region free of singularities ({\it i.e.}, with a
compactly generated Cauchy horizon) is incompatible with the
requirement that only normal matter be used ({\it i.e.}, that the
weak energy condition be satisfied).  The Tipler-Hawking
theorems, however, leave open the possibility of CTC formation if
a singularity appears (rendering the local region of spacetime
non-compact).  It is tempting to assume that any such CTC's will
be hidden behind an event horizon, but that has not been
proven.\fn{1}{Indeed, the cosmic censorship conjecture (on which
this assumption is based) has recently been brought into question
by numerical simulations \rf{6}.  Moreover, Ori \rf{7} has
recently argued that the singularities required by the
Tipler-Hawking theorems need not prevent the creation of CTC's.}
We review the Tipler-Hawking theorems in Appendix A.

Scientific interest in CTC's has been invigorated by Gott's
\rf{8,9} construction of an extraordinarily simple solution to
Einstein's equations that contains CTC's.  Gott's solution
describes two infinitely long parallel cosmic strings moving past
each other at high velocity.  The situation at early times is
portrayed in Fig.~1,
  \def\captiona{{\it A spacelike slice through the Gott
    spacetime.} Two parallel cosmic strings perpendicular to the
    page, represented by dots, move past each other at high
    velocity.  A deficit angle (shaded) is removed from the space
    around each string, with opposite sides identified at equal
    times.  (If the deficit angles were oriented in any direction
    other than along the motion of the string, the
    identifications would be at unequal times.) Note that no
    CTC's pass through this spacelike surface, as explained in
    the text.}%
  \figureinsert{1}{4in}{3in}{0.0in}{\captiona}
    {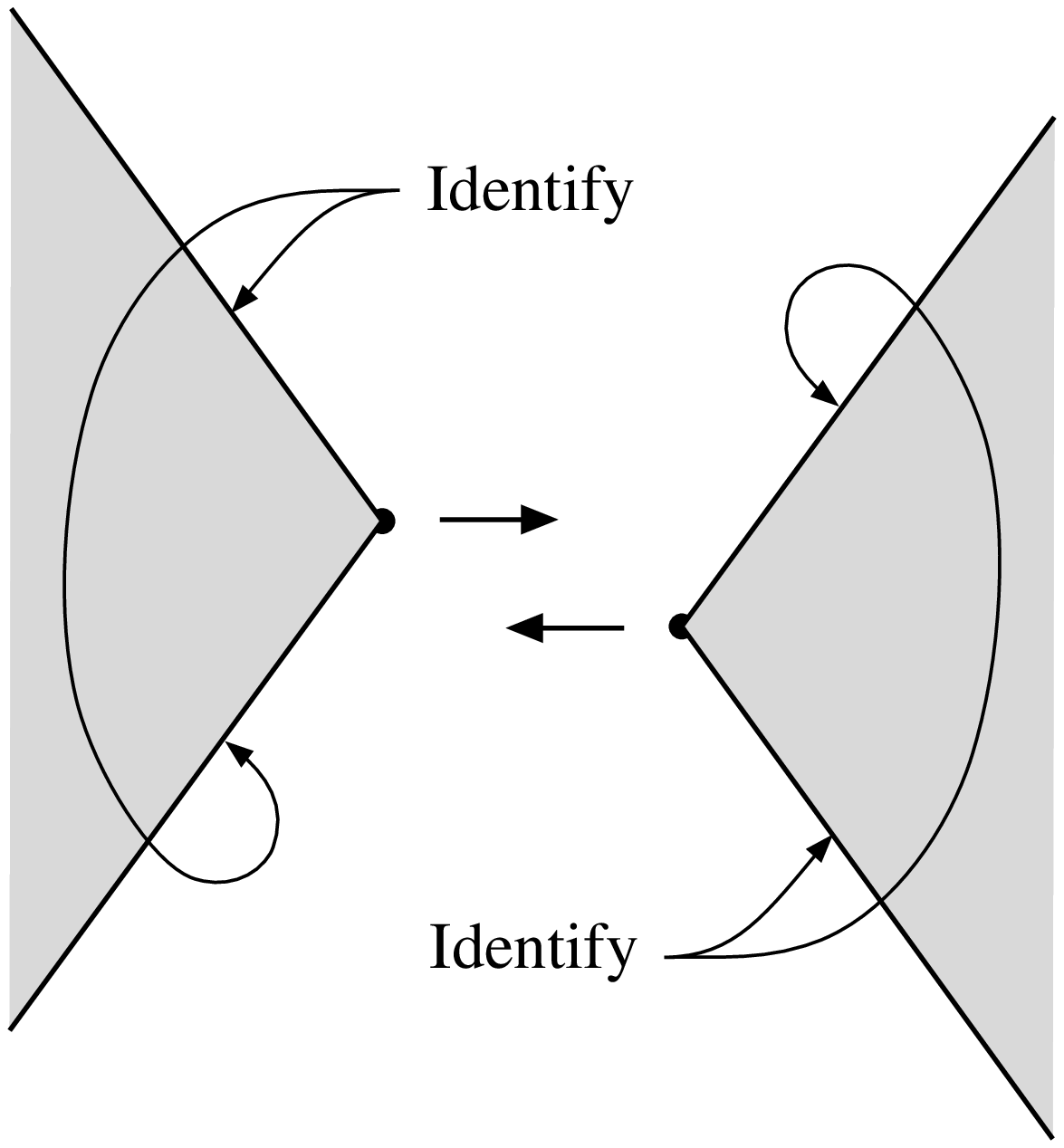 hoffset=-32 voffset=-384 hscale=63 vscale=63}%
which shows the two strings approaching each other.  Each string
is associated with a deficit angle removed from the space, which
we have oriented in a direction opposite to that of the motion of
the string.  Opposite sides of the excluded wedges are identified
at equal times.  Gott found that, as the strings approach each
other, it becomes possible to traverse a closed timelike curve
encircling the strings in the sense opposite to their motion.
The spacetime is topologically equivalent to Minkowski
space\fn{2}{In discussing the topology of the Gott universe, we
are treating the strings as objects with a small but nonzero
thickness.  They are nonsingular configurations, and are not
excised from the spacetime.} and free of singularities and event
horizons.

It is interesting to ask how the Gott solution is reconciled with
the Tipler-Hawking theorems.  Although cosmic strings have never
been observed, there is no reason to believe that they are
unrealistic forms of energy and momentum; certainly they satisfy
the weak energy condition, as required for the no-CTC theorems.
However, the strings in Gott's solution are infinitely long, so
the CTC's clearly do not arise in a local region.  Thus, in their
original form, the theorems of Tipler and Hawking have nothing to
say about the Gott time machine.

With further analysis, however, it can be seen that the infinite
length of the strings does not free the Gott spacetime from the
implications of arguments similar to those of Tipler and Hawking.
The relevance of the theorems can be established by exploiting a
special feature of the Gott spacetime: its equivalence to a
spacetime with point masses in (2+1) dimensions.  Any
(3+1)-dimensional spacetime populated solely by infinitely long
parallel cosmic strings is invariant under boosts and
translations along the direction of the strings, so the trivial
dependence of the metric on this direction can be ignored.  The
strings are then described as particles in (2+1) dimensions, and
the Gott time machine consists of two particles moving toward
each other at high speed.  Although the Tipler-Hawking theorems
were originally proven in the context of (3+1)-dimensional
general relativity, they may be extended to the (2+1)-dimensional
case, as we discuss in Appendix A\. The reconciliation of these
theorems with the Gott spacetime, therefore, involves issues more
subtle than the infinite length of the strings.

In an investigation of the causal structure of the Gott
spacetime, Cutler \rf{10} showed that it contains regions free of
CTC's, and constructed a complete spacelike hypersurface for
which there are no CTC's in the past.  A simpler example of such
a hypersurface is shown in Fig.~1; the past light cone of any
point on this equal-time surface extends through similar surfaces
arbitrarily far into the past, implying that no timelike curve
through such a point can be closed.  As the two particles
approach one another, each will ultimately collide with the
trailing deficit angle of the other, at which time this
coordinate system will fail; CTC's will then arise.  These CTC's,
however, do not originate in a compact region of spacetime; the
chronology horizon (the boundary of the region containing CTC's)
extends indefinitely in the past direction, although it never
intersects the hypersurface of Fig.~1.  Thus even when we
consider Gott's spacetime as a (2+1)-dimensional universe of
point particles, the CTC's are still not compactly generated.

Nevertheless, in order to explore how classical general
relativity complies with the Tipler-Hawking theorems, one can
imagine attempting to construct a Gott-like time machine.
Suppose, in a (2+1)-dimensional universe previously free of
CTC's, that two particles are accelerated toward each other in an
attempt to reach the velocity needed for a Gott time machine.
Since the consequences of this acceleration would be confined to
the future of the region in which it occurs, the chronology
horizon could not extend indefinitely toward the past, as it does
for the full Gott spacetime.  In analogous situations in (3+1)
dimensions, it is conceivable that the creation of CTC's is
permitted, and the theorems merely imply that singularities are
produced in the process. If so, the CTC's may or may not be
hidden by event horizons.  Alternatively, the creation of CTC's
might be strictly forbidden.  One could imagine, for example,
that any attempt to create a time machine would be thwarted by
energy loss due to gravitational radiation.  In (2+1) dimensions,
however, neither singularities nor gravitational radiation can
occur.  The theorems of Tipler and Hawking imply, however, that
some mechanism must prevent the formation of a Gott time machine
in these circumstances.  Because these (2+1)-dimensional systems
are exactly solvable, we will be able to study this mechanism in
detail.

The (2+1)-dimensional theory has been the object of extensive
investigation \rf{11,12}. It has been found that the metric in
vacuum is necessarily flat,\fn{3}{We will always assume that the
cosmological constant vanishes.} while in the presence of a
single particle with mass $M$, the external metric is that of
Minkowski space from which a wedge of angle $\alpha=8\pi GM$ has
been removed and opposite sides have been identified ($G$ is
Newton's constant).  Solutions with several static particles are
easily constructed by joining several one-particle solutions, in
which case the space has a net deficit angle given by the sum of
the deficit angles of the constituent particles. If the total
deficit angle of a space with static particles exceeds $2\pi$,
then the spatial sections of the spacetime must be closed
\rf{13}; the topology is $S^2$, and the total deficit angle is
necessarily exactly $4\pi$. By joining appropriately boosted
single-particle spacetimes, the exact solutions with moving
particles \rf{12} and nontrivial decays and scatterings \rf{14}
may be constructed.

In their seminal paper on (2+1)-dimensional gravity, Deser,
Jackiw, and 't~Hooft \rf{12} noted that a spinning point particle
would give rise to CTC's.  They added, however, that ``such
closed timelike contours are not possible in a space with $n$
moving spinless particles, where angular momentum is purely
orbital.'' When stated without qualification, this sentence is
apparently contradicted by the existence of the Gott solution.
What is true, as implied by the (2+1)-dimensional version of the
Tipler-Hawking theorems, is that CTC's cannot be created from
scratch in a local region.  The mechanism by which these theorems
are enforced will be the main subject of this paper.

The energy and momentum of a collection of particles can be
conveniently characterized by the Lorentz transformation that a
vector undergoes upon being parallel transported around the
system \rf{12,15}.  (In the presence of CTC's we must be careful
in choosing the loop which surrounds the particles, as we explain
below.) The Lorentz transformation belongs to the
(2+1)-dimensional Lorentz group, SO(2,1).  Deser, Jackiw, and
't~Hooft \rf{16} have shown that the group element corresponding
to the Gott time machine is boostlike, {\it i.e.}, equivalent
under similarity transformation to a pure boost.  Each element of
SO(2,1) can be identified with a 3-vector, which we will refer to
as the energy-momentum vector of the system. The energy-momentum
vector for a single particle is timelike, while that of the Gott
two-particle system is spacelike (tachyonic), despite the fact
that each particle is moving slower than $c$.

The possibility of timelike momenta combining to form a spacelike
momentum arises because the momentum of a system of particles is
not the sum of the individual momenta.  In (2+1)-dimensional
gravity, unlike special relativity, the composition law for
energy-momentum vectors is nonlinear. Hence it becomes possible
for a system composed of ordinary, subluminal matter to have the
same energy-momentum vector as a tachyonic particle.\fn{4}{The
energy-momentum vector, however, does not tell the whole story.
The external spacetime associated with a fundamental tachyon is
different from that associated with a Gott time machine, as we
will make explicit in \sec II\.}

One might guess that this discovery reveals why it is impossible
to build a Gott time machine from slowly-moving particles:
because the momentum is tachyonic, and we can exclude tachyons as
unphysical.  In fact, we will argue that the momentum of Gott's
time machine is not unphysical: it is possible, given sufficient
energy, to produce a Gott pair from the evolution of initially
static particles, even in a theory that does not contain
fundamental tachyons.

There are, nevertheless, insuperable barriers to creating a Gott
time machine, by which we mean a system of particles with
spacelike total momentum that leads to the creation of CTC's. In
an open universe, the obstacle is that there is never sufficient
energy.  In our previous paper \rf{14} we examined a specific
scenario --- the decay of two initially static particles in an
open universe --- and showed that the offspring particles could
never move fast enough to make a Gott time machine.  In this
paper we generalize the result, by examining the evolution
determined by arbitrary initial data specified on an edgeless
spacelike surface $S$.  A necessary condition for a Gott time
machine to evolve from such data is that a subsystem of particles
in the domain of dependence of $S$ (or on the boundary of the
domain of dependence) have a spacelike momentum.  We show that if
$S$ has the topology of $\IR^2$ and the total momentum passing
through $S$ is timelike, then there is insufficient energy for a
spacelike subsystem to arise. In this sense, a Gott time machine
cannot be created in an open universe with timelike total
momentum.

As the key step in our demonstration, we associate with every
collection of particles an element of the universal covering
group of SO(2,1), and show that an element corresponding to an
open hypersurface with timelike momentum is never the product of
an element representing the Gott time machine and any number of
elements representing massive particles.  The proof can be
constructed by algebraic manipulation; however, an elegant
geometric demonstration is achieved by introducing an invariant
metric on the parameter space of the group, in which case the
group manifold becomes (2+1)-dimensional anti-de Sitter
space.\fn{5}{We are very grateful to Don Page and Alex Lyons, who
pointed out to us the relationship between the (2+1)-dimensional
Lorentz group and anti-de Sitter space.}

In a closed universe, this result no longer holds --- we have
found that it is possible to construct a spacetime in which a
pair of particles with tachyonic momentum is created from
initially static conditions ({\it i.e.,} by the decay of massive
particles).  However, 't~Hooft \rf{17} has shown that causal
disaster is avoided in this case, since the universe shrinks to
zero volume before any CTC's can arise.\fn{6}{In a note added to
our previous paper \rf{14} we erroneously claimed that CTC's
would arise.  We thank G. 't~Hooft for informing us of our
mistake.} He goes on to argue that this phenomenon will always
result from an attempt to build a time machine in a closed
universe. Therefore, neither closed nor open universes in (2+1)
dimensions can evolve Gott time machines from initial conditions
with slowly-moving particles.  These cases provide concrete
illustrations of the mechanisms that enforce the theorems of
Tipler and Hawking.

\head{II. ENERGY AND MOMENTUM}

\subhead{A. Overview}

This section is devoted to a detailed discussion of the use of
holonomy --- the Lorentz transformation associated with parallel
transport of a spacetime vector around a closed loop --- to
quantify the energy and momentum of gravitating point particles
in (2+1) dimensions.  With a suitable choice of coordinate
system, the holonomy is a simple function of the velocities and
deficit angles of the particles enclosed by the loop.

The parameter space of SO(2,1), the group of Lorentz
transformations in (2+1) dimensions, can be given an invariant
metric, establishing a correspondence between the group manifold
and (2+1)-dimensional anti-de~Sitter space.  We find that there
are inequalities that relate the momentum of a system of
particles to the momenta of the constituents, and that these
inequalities can be expressed compactly by referring to the
causal structure of the anti-de Sitter metric. In the process, we
find it useful to extend SO(2,1) to its universal covering group,
in which rotations are not identified modulo $2\pi$, and which
can therefore distinguish between (for example) a universe with
no matter and a universe with total deficit angle $2\pi$ or
$4\pi$.

We wish to comment that many of the tools we use are standard
results in the theory of Lie groups and symmetric spaces \rf{18}.
Any Lie algebra has a natural metric, the Cartan-Killing form,
given in components by
  $$g_{\mu\nu}=c^\lambda{}_{\mu\sigma} c^\sigma{}_{\nu\lambda} \ ,
  \eqno(1)$$
where the $c^{\alpha}{}_{\beta\gamma}$ are the structure
constants. For semisimple groups, this metric can be uniquely
extended to a left- and right-invariant metric on the entire
group manifold. The resulting space will be maximally symmetric,
and paths from the identity defined by $T(\lambda)=\exp(-i\lambda
X)$, where $X$ is an element of the Lie algebra, will be
geodesics of this metric.  However, it is straightforward for us
to derive these results for the case at hand, which we will do
for clarity.  (A similar discussion of the universal cover of
SO(2,1) in a different context can be found in \Ref{19}.  A more
mathematical discussion can be found in \Ref{20}.)

\subhead{B. Holonomy}

We begin by recalling the characterization of energy and momentum
in (2+1)-dimensional gravity \rf{12,15}.  In any number of
dimensions, spacetime curvature can be characterized by the
holonomy transformation that describes the result of parallel
transporting a vector around a closed loop.  In (2+1) dimensions
this technique is especially convenient, since the holonomy of a
(contractible) closed loop can be thought of as reflecting the
energy and momentum of the matter passing through the loop.
Further, since space is flat outside sources, any two loops that
can be deformed into each other without crossing any particles
will yield the same transformation.  (The holonomy will, of
course, depend on a choice of frame at the base point of the
loop. If the choice of frame is varied, the holonomy will change
by a similarity transformation.)

In a general manifold, the holonomy around a loop is a
path-ordered exponential of the connection.  In the case at hand,
however, the flatness of spacetime in vacuum affords a
considerable simplification.  As an example, consider a single
particle that is stationary in a Minkowskian coordinate system
from which a wedge of deficit angle $\alpha$ has been removed,
with opposite sides identified. A vector parallel transported in
a counterclockwise loop around the particle has constant
components until it crosses the identified edges, where it
undergoes a rotation by $\alpha$.  The holonomy of this loop is
therefore a counterclockwise rotation matrix $R(\alpha)$.  For a
particle in a similar coordinate system but moving with velocity
$\vec v$, the appropriate transformation can be obtained by
boosting to the particle's rest frame, rotating, and boosting
back.  Thus, we associate with the moving particle a matrix
  $$T=B(\vecxi)R(\alpha)B^{-1}(\vecxi) \ ,\eqno(2)$$
where $\vecxi =\hat v \tanh^{-1}|\svec v|$ is the rapidity of the
particle and $B(\vecxi)$ is a boost bringing the rest vector to
the velocity of the particle.

If there are several particles moving with respect to one
another, then the holonomy around any loop can be computed by
deforming the loop so that it goes around the particles one at a
time.  The holonomy is then a product of matrices, each of which
has the form of \eq (2).

The discussion of multiparticle systems is particularly simple in
the context of a spacetime (or region of spacetime) free of
CTC's, since in this case we can construct a foliation into
spacelike surfaces.  Specifically, the momentum of a system of
particles contained in a bounded, simply connected region of a
spacelike surface is characterized by the holonomy of the
(counterclockwise) loop that forms the boundary of the region.
(We consider the particle worldlines to be part of the manifold,
so such worldlines do not render a region multiply connected.)
The base point of the holonomy can be thought of as the position
of the observer, and defines the coordinates in which the
holonomy transformation is measured. The relation between this
holonomy and the properties of the individual particles is easily
seen by continuously deforming the loop into one that encircles
the particles one at a time, as shown in Fig.~2.
  \def\captionb{{\it The deformation of a loop into a succession
    of single particle loops.} Part (a) shows three particles
    contained in a shaded region, bounded by the counterclockwise
    loop $C$ with base point $Q$.  To deform the loop, first draw
    non-intersecting paths $P_i$ connecting each particle to the
    base point, as shown by the dashed lines in (b).  (These
    paths are arbitrary, but in a connected region they can
    always be constructed.) Then deform the loop so that the area
    inside shrinks, continuing until all parts of the loop come
    into contact with either the particles or the paths $P_i$.
    The result is a path $C'$, as shown in (c), which encircles
    the particles one at a time.}%
  \figureinsert{2}{1.72in}{3.75in}{0.0in}{\captionb}
    {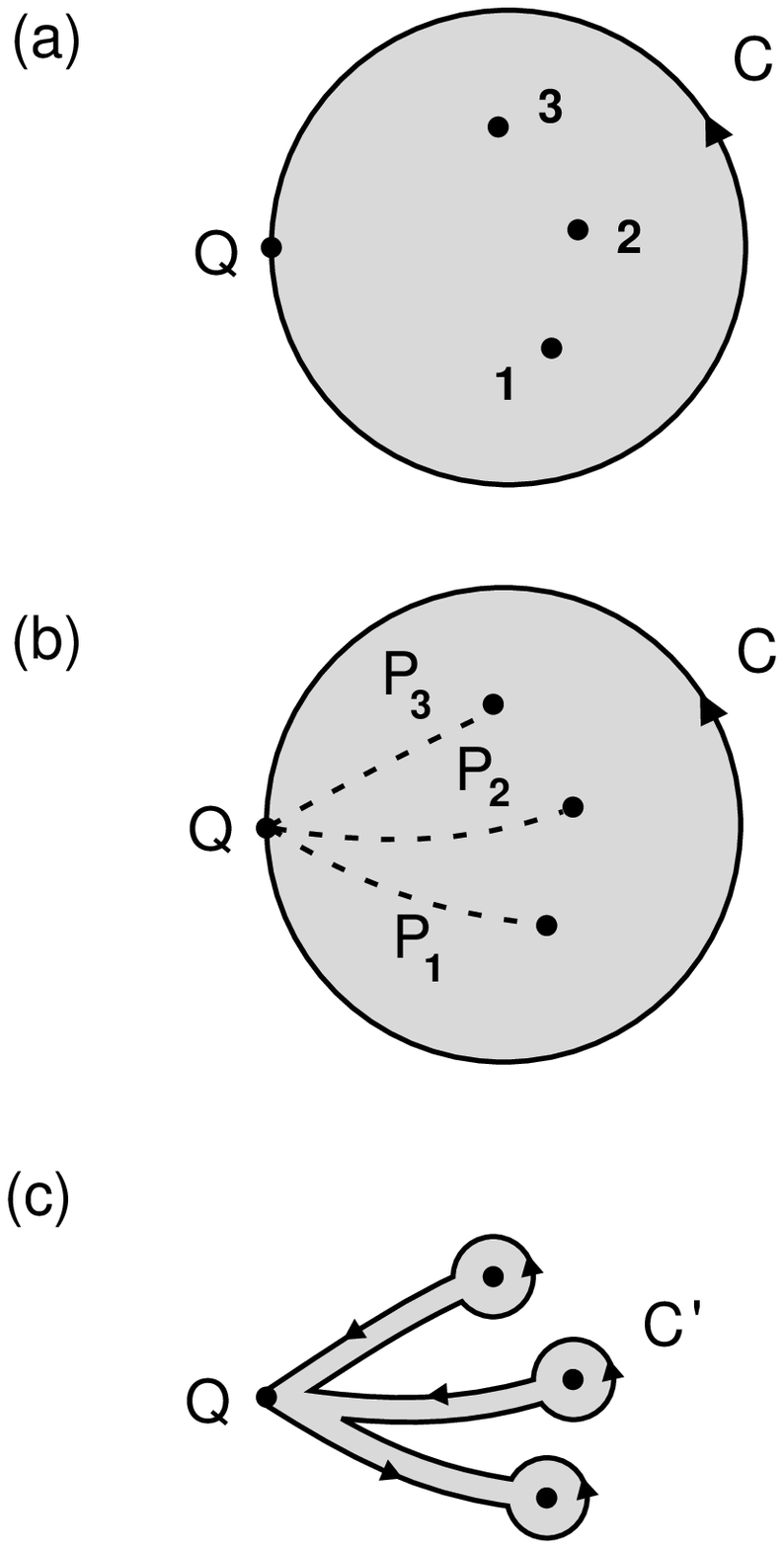 hoffset=-97 voffset=-386 hscale=57 vscale=57}%
Such a deformation can always be constructed by first choosing
non-intersecting paths $P_i$ to connect each particle to the base
point $Q$.  (In a coordinate system adapted to this picture, the
deficit angles would not cross the paths $P_i$ but otherwise may
extend in any direction.) The original loop $C$ is then deformed
to the loop $C'$ by continuously shrinking the area inside the
loop, continuing until all parts of the loop come into contact
with either the particles or the paths $P_i$. The parallel
transport of a vector around the path $C'$ is thus the product of
the parallel transport transformations of the loops around the
individual particles, so
  $$T_{tot}=T_N T_{N-1}\ldots T_1\ ,\eqno(3)$$
where the particles are enclosed in the order $1,2,\ldots N$.
Here each $T_i$ has the form of \eq (2), where the velocity of
the $i$th particle is determined by parallel transporting the
velocity vector to the base point $Q$ along the path $P_i$.

As a system of $N$ particles evolves via decay or scattering into
a system of $M$ particles, a loop around the system at one time
can be deformed to a loop around the system at a later time ({\it
i.e.}, on a subsequent spacelike surface). Since the deformation
carries the loop only through regions of flat spacetime, the
resulting transformation matrix is not changed. Therefore,
conservation of energy and momentum is expressed as the equality
of Lorentz transformations at different times:
  $$T_N T_{N-1}\ldots T_1=T^\prime_M T^\prime_{M-1}\ldots
     T^\prime_1 \ . \eqno(4)$$
This will hold regardless of the orientations of the paths $P_i$,
as long as the base points of the two loops lie in the same
Minkowskian coordinate patch. This rule was used in \Ref{14} in
constructing the spacetime for one particle decaying into two.

Later in this paper we will be concerned with the relation
between a system of particles and a subset of the system.  The
subsystem is defined by specifying the region of the spacelike
surface that it occupies.  To obtain a holonomy that can be
compared with the holonomy of the full system, the base point $Q$
of the full holonomy should also lie on the boundary of the
subregion, and should be taken as the base point for its
holonomy. Note that the specification of the region contains more
information than a simple listing of the particles in the
subsystem; Fig.~3 shows an example of two distinct regions that
contain exactly the same particles.
  \def\captionc{{\it Two distinct regions containing the same
    particles.} The shaded regions in parts (a) and (b) each
    contain the same particles, 1 and 2.  Nonetheless, the
    boundary loops $C_a$ and $C_b$ cannot be continuously
    deformed into one another without crossing particle 3, and
    therefore the holonomies of the two loops will be unequal.}%
  \figureinsert{3}{4.03in}{2.00in}{0.0in}{\captionc}
    {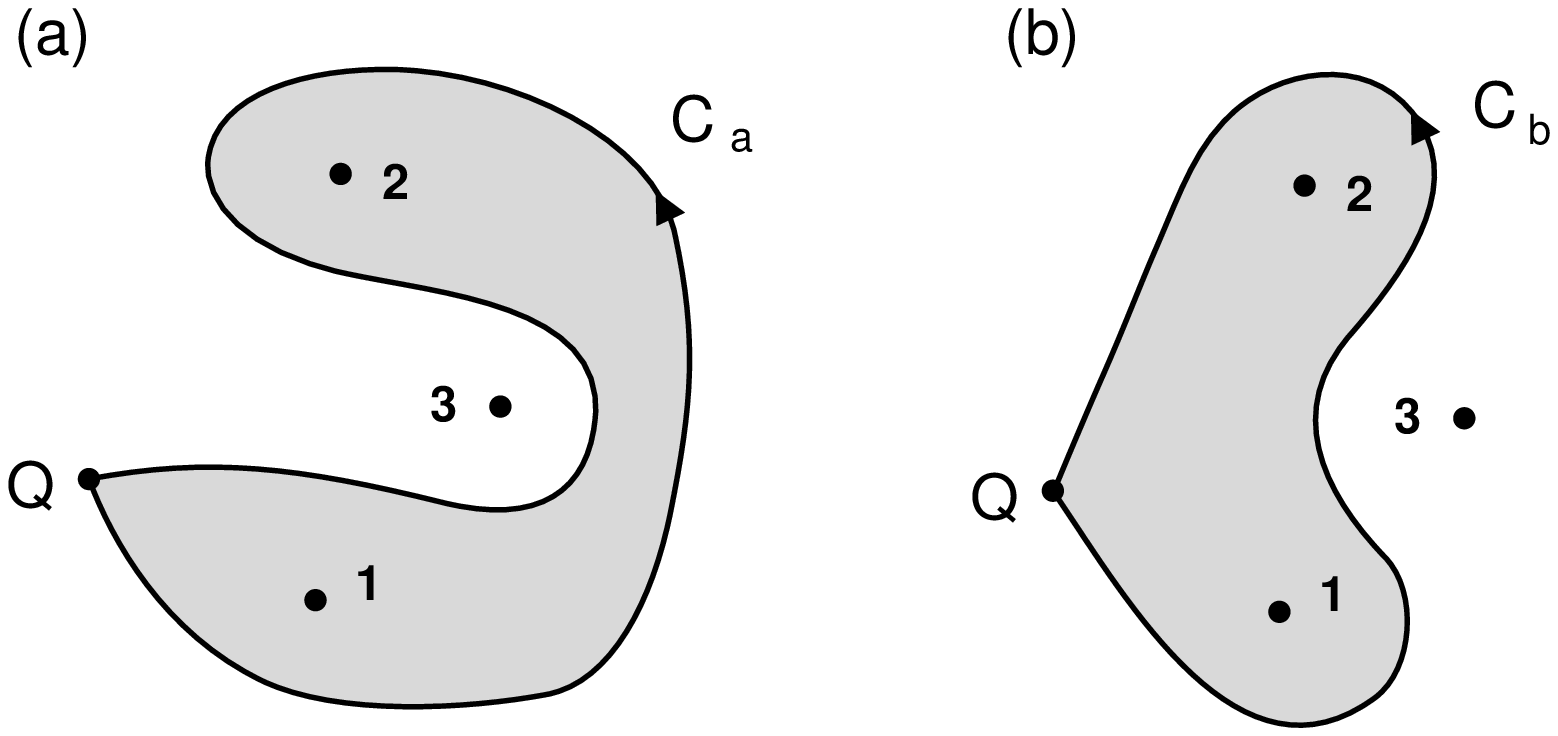 hoffset=-35 voffset=-334 hscale=61 vscale=61}%
To understand the relevance of the extra information, recall that
the holonomy defines the energy and momentum of the subsystem,
and therefore must include a specification of the relative
velocity of the two particles.  The velocity of particle 2 as
seen by particle 1, however, depends on the path used for the
observations.  That is, the velocities can be compared only by
parallel transporting one to the other, a path-dependent process.

The generic form of a system and subsystem are shown in Fig.~4a.
  \def\captiond{{\it The relation between a subsystem and the
    entire system.} Part (a) shows a generic system and
    subsystem, bounded by loops $C_{tot}$ and $C_{sub}$,
    respectively.  The loop $C_{tot}$ can be deformed, as shown
    in (b), by extending it along the path of $C_{sub}$ and then
    back again.  Part (c) shows how the new loop, $C_{tot}'$, can
    be viewed as the concatenation of two loops: $C_{sub}$, which
    surrounds the subsystem, and $C_{else}$, which surrounds the
    remainder of the system.}%
  \figureinsert{4}{4.0in}{4.06in}{0.0in}{\captiond}
    {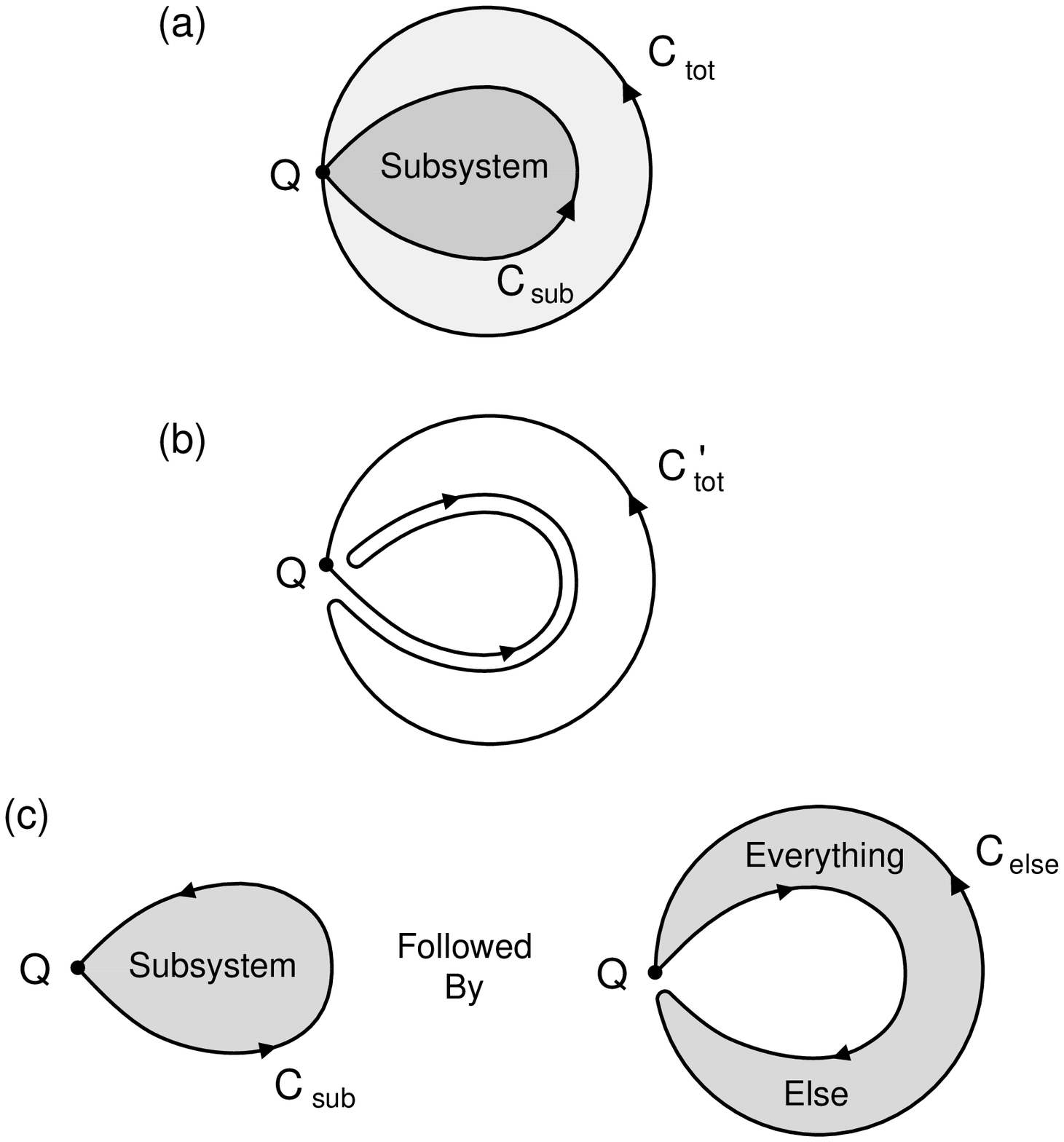 hoffset=-32 voffset=-372 hscale=58 vscale=58}%
The loop around the system is called $C_{tot}$, and the loop
around the subsystem is called $C_{sub}$. To relate the total
holonomy $T_{tot}$ to the holonomy of the subsystem $T_{sub}$,
first deform $C_{tot}$ to $C_{tot}'$, as shown in Fig.~4b.  The
deformation consists of adding a detour which runs along
$C_{sub}$, and then returns along the same path.  Then note that
the loop $C_{tot}'$ can be viewed as a sequence of two loops, as
shown in Fig.~4c.  The first is a loop $C_{sub}$, surrounding the
subsystem, and the second is a loop $C_{else}$ that surrounds the
remainder of the system.  The holonomy for the full system can
then be written as
  $$T_{tot} = T_{else}T_{sub} \ .\eqno(5)$$
$C_{else}$, however, is a loop surrounding a simply connected
region, so it can be decomposed into its single particle
contributions as in \eq (3). Thus, $T_{tot}$ can always be
written as
  $$T_{tot} = T_N T_{N-1}\ldots T_1 T_{sub} \ ,\eqno(6)$$
where the $T_i$ denote matrices of the form of \eq (2), for each
particle not part of the subsystem.  This equation implies that
no matter how the subsystem is chosen, the complete system can
always be separated cleanly into the subsystem plus other
particles.  \eq (6) will be used later to prove an inequality
relating the energy-momentum of a subsystem to that of the system
in which it is contained.

We turn now to a specific representation of the holonomy
transformations $T$ by $2 \times 2$ matrices.  The standard basis
for the Lie algebra of SO(2,1) consists of the rotation generator
$J$ and two boost generators $K_1$ and $K_2$, with the
commutation relations
  $$\eqalign{ &[K_1,K_2] = -iJ\cr
              &[J,K_1] = i K_2\cr
              &[J,K_2] = - i K_1 \ .\cr}\eqno(7)$$
Following the conventions used in \Ref{14}, we take $J = {1 \over
2} \sigma_3$ and $K_i = {i \over 2} \sigma_i$, where the
$\sigma$'s are the standard Pauli matrices. When exponentiated,
these $2 \times 2$ matrices generate the group SU(1,1), which is
a double cover of SO(2,1). Since our ultimate concern will be the
universal cover common to SO(2,1) and SU(1,1), for convenience we
will work with SU(1,1) in what follows.

The generators $J$ and $K_i$ are components of an antisymmetric
Lorentz tensor $M_{\mu\nu}$, with $J = M_{12}$ and $K_i =
M_{i0}$.  Since we are working in three spacetime dimensions,
however, we can define a set of 3-vector group generators by
using the Levi-Civita tensor:
  $$\J_\mu = {1 \over 2}\epsilon_\mu{}^{\lambda\sigma}
     M_{\lambda\sigma} \ ,\eqno(8)$$
where $\epsilon_{012} \equiv 1$, and indices are raised and
lowered by the Lorentz metric $\eta_{\mu \nu} \equiv \diag
[-1,1,1]$.  Explicitly, $\J_0 = J$, $\J_1 = -K_2$, and $\J_2 =
K_1$.

A rotation by angle $\alpha$ is given by
  $$R(\alpha)= e^{- i \alpha J} = \gmatrix{e^{-i\alpha/2}&0\cr 0&
     e^{i\alpha/2}\cr} \eqno(9)$$
and a boost is given by
  $$B(\vecxi)=e^{- i \vecxi \cdot \vec K} =
     \gmatrix{\cosh{\xi\over 2}&e^{-i\phi}\sinh {\xi\over 2}\cr
     e^{i\phi}\sinh{\xi\over 2} & \cosh{\xi\over 2}\cr}
     ,\eqno(10)$$
where $\xi=|\vecxi |$ is the magnitude of the rapidity and $\phi$
is its polar angle.  The matrix $T$ associated with a single
particle is found by evaluating \eq (2):
  $$T = \gmatrix{ \sqrt{1 + p^2} e^{-i \alpha' /2} & i p e^{- i
     \phi}\cr - i p e^{i \phi} & \sqrt{1 + p^2} e^{i \alpha'
     /2}\cr} ,\eqno(11)$$
where
  $$p \equiv \sinh \xi \sin {\alpha \over 2} \eqno(12)$$
is a measure of the momentum of the particle, and $\alpha^\prime$
is defined by
  $$\tan {\alpha' \over 2} = \cosh \xi \tan {\alpha \over 2} \
     .\eqno(13)$$
Note that $\alpha^\prime$ may be thought of as a boosted deficit
angle, when the excised wedge is taken to lie along the direction
of motion (as in Fig.~1).

\subhead{C. The Metric on SU(1,1)}

We now turn to the geometry of the parameter space of SU(1,1).
The space of $2 \times 2$ matrices is spanned by the group
generators $\J_\mu$ ($\mu=0,1,2$) and the identity matrix, so an
arbitrary $2 \times 2$ matrix can be written as
  $$T = w - 2 i \chi^\mu \J_\mu =
     \gmatrix{w - i t & y + i x\cr
             y - i x & w + i t\cr } ,\eqno(14)$$
where $\chi^\mu \equiv (t,x,y)$, and $w$, $t$, $x$, and $y$ are
complex. SU(1,1) consists of those matrices $T$ satisfying
  $$\det T=+1 \eqno(15a)$$
and
  $$T^\dagger \eta T= \eta \ ,\eqno(15b)$$
where
  $$\eta = \gmatrix{1&0\cr 0&-1\cr} .\eqno(16)$$
It is shown in Appendix B that these conditions are obeyed if and
only if $(t,x,y,w)$ are real numbers satisfying
  $$-t^2 + x^2 + y^2 - w^2 = -1 \ . \eqno(17)$$
The form of this equation suggests that we consider the
three-dimensional space indicated by \eq (15) to be embedded in a
four-dimensional space with metric
  $$ds^2=-dt^2 + dx^2 + dy^2 - dw^2 \ .\eqno(18)$$
It is natural to take the metric on the parameter space of
SU(1,1) to be the metric induced by this embedding.  Aficionados
of de~Sitter spaces will recognize this as (2+1)-dimensional
anti-de~Sitter space (see \Ref{3}).  The group SO(2,2) which
leaves this metric invariant will map the submanifold defined by
\eq (17) into itself, so the embedded three-manifold is maximally
symmetric.  Furthermore, it will be shown in Appendix B that the
metric is group invariant--- that is, either left- or
right-multiplication by an element of SU(1,1) is an isometry of
the metric.

To put coordinates on SU(1,1), note that we can decompose any
element into a boost times a rotation:
  $$\eqalign{T&=B(\vec\zeta)R(\theta)\cr
  &=\gmatrix{e^{-i\theta/2}\cosh{\zeta\over 2}&
  e^{i(\theta/2-\delta)}\sinh{\zeta\over 2}\cr
  e^{i(-\theta/2+\delta)}\sinh{\zeta\over 2}&
  e^{i\theta/2}\cosh{\zeta\over 2}\cr} ,\cr}\eqno(19)$$
where $\zeta$ is the magnitude of the boost, $\delta$ is its
polar angle, and $\theta$ is the angle of rotation.  (Note that
this is a different parameterization than the variables
($\alpha$, $\xi$, $\phi$) used in \eqs (11-13).) Comparing
\eqpar{19} to \eq (14) and defining $\psi=2\delta-\theta$, we
obtain
  $$\eqalign{t &= \sin {\theta \over 2} \,
     \cosh{\zeta \over 2} \cr
   x &= \sin {\psi \over 2} \, \sinh {\zeta \over
     2} \cr
   y &= \cos {\psi \over 2} \, \sinh {\zeta \over
     2} \cr
   w &= \cos {\theta \over 2} \, \cosh{\zeta \over
     2}  \ . \cr}\eqno(20)$$
The anti-de~Sitter metric on SU(1,1) is the metric induced by \eq
(18) on the submanifold defined by \eq (17). In the coordinates
$(\theta, \zeta,\psi)$, it is obtained by plugging the
transformations \eqpar{20} into \eq (18), yielding
  $$ds^2=-{1\over 4}\cosh^2{\zeta\over 2}\ d\theta^2+{1\over
     4}d\zeta^2 +{1\over 4}\sinh^2{\zeta\over 2}\
     d\psi^2 \ .\eqno(21)$$
Thus $\theta$ acts as a timelike coordinate.

\subhead{D. The Energy-Momentum Vector}

Elements of SU(1,1) sufficiently close to the identity may be
written as exponentials of elements of the Lie algebra:
  $$T=\exp{(-i\phi^\mu{\J}_\mu)} \ .\eqno(22)$$
Since $\phi^\mu$ describes a tangent vector at the identity of
SU(1,1), we can compute its norm using the metric defined above.
One could use \eq (21), but it is easier to expand $T$ to first
order in $\phi^\mu$, and then to compare with \eq (14) to
determine the parameters $(dt,dx,dy,dw)$ needed in \eq (18).  The
result is
  $$|\phi|^2 = {1 \over 4} \eta_{\mu\nu} \, \phi^\mu \, \phi^\nu
     \ .  \eqno(23)$$
Not surprisingly, the group invariant metric in the vicinity of
the identity coincides, up to a factor, with the usual
(2+1)-dimensional Minkowski metric $\eta_{\mu\nu}$.  If
$|\phi|^2<0$ we will call the vector ``timelike,'' keeping in
mind the distinction between the timelike direction in the
spacetime manifold and that on SU(1,1). Note that $|\phi|$ is
actually the length of the curve defined by $T(\lambda)=\exp{(-i
\lambda \phi^\mu{\J}_\mu)}$, where $\lambda$ varies from 0 to 1,
as can be seen by first calculating the length of the segment
from $\lambda$ to $\lambda + d \lambda$.

To understand the properties of $\phi^\mu$, let us write the
holonomy of a loop around a single particle in the form
$T=\exp(-i\phi^\mu{\J}_\mu)$. Under a Lorentz transformation $L$
in the physical spacetime, the group element
$\exp(-i\phi^\mu{\J}_\mu)$ transforms as
  $$L\exp{(-i\phi^\mu{\J}_\mu)}L^{-1}=
  \exp(-i\Lambda^\mu{}_\nu\phi^\nu{\J}_\mu) \ ,\eqno(24)$$
where $L$ is an element of the $2\times 2$ representation of
SU(1,1) and $\Lambda$ is the corresponding matrix in the $3\times
3$ (adjoint) representation.  In the rest frame of a single
particle $T$ is a pure rotation and $\phi^\mu$ is equal to
$(\alpha,0,0)=(8\pi GM,0,0)$.  Using \eq (24), one sees that in
an arbitrary frame $\phi^\mu=8\pi G(\gamma M,\gamma M \svec v)$,
in agreement (up to a factor) with the energy-momentum vector of
special relativity.  It follows immediately that any massive
particle (moving slower than the speed of light) will be
associated with a $\phi^\mu$ that is timelike.  (By convention,
the holonomy of a counterclockwise loop corresponds to a
future-directed energy-momentum vector.) If several particles are
combined, however, then $\phi^\mu_{tot}$ is not the sum of the
individual momenta; rather, from \eq (3),
  $$\exp(-i\phi^\mu_{tot}{\J}_\mu)=\exp(-i\phi^\mu_N{\J}_\mu)
  \exp(-i\phi^\mu_{N-1}{\J}_\mu)\ldots
  \exp(-i\phi^\mu_1{\J}_\mu) \ .\eqno(25)$$
In the $G \to 0$ limit, on the other hand, each exponential can
be expanded to lowest order, and one finds that $\phi^\mu_{tot}$
approaches the sum of the individual $\phi^\mu_i$.  Since
$\phi^\mu_{tot}$ is a conserved Lorentz 3-vector which approaches
the ordinary special relativistic energy-momentum vector in the
$G \to 0$ limit, we will call it the energy-momentum vector.
However, it should be recognized that it is a somewhat
unconventional energy-momentum vector in at least two respects.
First, and more importantly for our purposes, the energy-momentum
vector for a group of particles is not the sum of the individual
momenta.  Second, it is constructed from the Lie algebra of the
Lorentz group, rather than the tangent space of the spacetime.
Nonetheless, in 2+1 dimensions the Lorentz generators can be
rearranged to form a vector, as in \eq (8).  The expansion of a
group element in terms of these generators gives rise to a
three-tuple $\phi^\mu$, which transforms according to \eq (24) as
a vector in the tangent space of the spacetime.  The tangent
space is constructed at the location of the base point of the
closed loop used for parallel transport, which might be thought
of as the location of the observer.\fn{7}{If multiple observers
are stretched along the loop, or along any deformation of the
loop that intersects no other particles, then the observers will
agree on the energy-momentum vector in the following sense: if
the vector measured by one observer is parallel transported along
the loop to another observer, it will agree with the vector
measured by the second observer.}

For some purposes it will be convenient to write $\phi^\mu$ as a
parameter $\lambda$ times a normalized vector $n^\mu$; if
$\phi^\mu$ is timelike $|n|^2=-1$, and for $\phi^\mu$ spacelike
$|n|^2=+1$.  For $n^\mu$ timelike, the explicit form for $T$ is
  $$\eqalign{e^{-i\lambda n^\mu{\J}_\mu} &=\cos{\lambda\over
     2}- 2in^\mu{\J}_\mu\sin{\lambda\over 2}\cr
   &=\gmatrix{
     \cos{\lambda\over 2}-in^0\sin{\lambda\over 2}&
     (n^2+in^1) \sin{\lambda\over 2}\cr
    (n^2-in^1)\sin{\lambda\over 2}&
     \cos{\lambda \over 2}+in^0\sin{\lambda\over 2}\cr} ,\cr}
     \eqno(26)$$
and for $n^\mu$ spacelike we obtain
  $$\eqalign{e^{-i\lambda n^\mu{\J}_\mu} &=\cosh{\lambda\over
     2}- 2in^\mu{\J}_\mu\sinh{\lambda\over 2}\cr
   &=\gmatrix{
     \cosh{\lambda\over 2}-in^0\sinh{\lambda\over 2}&
     (n^2+in^1) \sinh{\lambda\over 2}\cr
    (n^2-in^1) \sinh{\lambda\over 2}&
     \cosh{\lambda \over 2}+in^0\sinh{\lambda\over 2}\cr} .\cr}
     \eqno(27)$$
Taking the trace of these two equations, we can see that a
general matrix obtained by exponentiation satisfies
  $${1\over 2}\Tr \left[ e^{-i\lambda n^\mu{\J}_\mu}\right] =
  \cases{\cos{\lambda\over 2} & for $n^\mu$ timelike ,\cr
  \cosh{\lambda\over 2} & for $n^\mu$ spacelike .\cr}\eqno(28)$$
It follows that
  $${1\over 2}\Tr \left[ e^{-i\lambda n^\mu{\J}_\mu}\right]
     \ge -1 \eqno(29)$$
for all cases.

{}From \eq (29), it is easy to show that the SU(1,1) matrix
corresponding to the Gott time machine is not the exponential of
any generator. For simplicity, we take a configuration where the
two particles approaching each other (as in Fig.~1) each have
rest frame deficit angle $\alpha$ and rapidity $\xi$, with
$\phi_1=\pi$, $\phi_2=0$.  Then we can use \eqs (3) and (11) to
write $T_G=T_2T_1$ as
  $$T_G = \gmatrix{
     (1+p^2) e^{-i \alpha'} - p^2 & - 2 p \sqrt{1 + p^2} \sin
          {\alpha' \over 2} \cr
     - 2 p \sqrt{1 + p^2} \sin {\alpha' \over 2} & (1+p^2) e^{i
          \alpha'} - p^2 \cr} ,\eqno(30)$$
where $p$ and $\alpha'$ are given by \eqs (12) and (13).  The
trace is then given by
  $$\eqalign{{1\over 2}\Tr T_G &= (1+p^2) \cos \alpha' - p^2 \cr
   &= 1-2\cosh^2\xi \, \sin^2{\alpha\over 2}\ .\cr}\eqno(31)$$
The condition that such a configuration contain closed timelike
curves is \rf{8}
  $$\cosh\xi \, \sin{\alpha\over 2} >1 \ ,\eqno(32)$$
or ${1\over 2}\Tr T_G<-1$.  Thus, from \eq (29) it follows that
$T_G$ cannot be written as an exponential.

However, SU(1,1) is a double cover of the Lorentz group, so the
matrices $\pm T_G$ correspond to the same element of SO(2,1).  We
will see in the next section that $-T_G$ can be written as an
exponential.  Since ${1\over 2}\Tr (-T_G) >1$, \eq (28) implies
that it is the exponential of a spacelike generator.  The
corresponding element of SO(2,1) can be obtained by
exponentiating the corresponding generator, and thus the element
of SO(2,1) associated with a Gott pair is spacelike or tachyonic
(equivalent under similarity transformation to a pure boost).
This is the sense in which we say that the Gott time machine has
tachyonic momentum \rf{16,14}, even though $T_G\in$ SU(1,1) is
not equivalent to the exponential of a spacelike generator ---
parallel transport of a spinor around a single tachyonic particle
is not equivalent to parallel transport around the Gott
two-particle system, although parallel transport of an SO(2,1)
vector does not distinguish between the two cases.

\subhead{E. Constraints on the Energy-Momentum of Systems}

We now return to the anti-de Sitter geometry of SU(1,1). For
fixed $n^\mu$, we may consider \eqs (26) and (27) as defining
curves parameterized by $\lambda$.  The crucial observation is
that these curves are geodesics in the metric \eqpar{21}, which
can be checked directly. For example, by comparing \eqs (19) and
(26), one sees that the curve defined by \eqpar{26} is equivalent
to
  $$\eqalign{\theta&=2\tan^{-1}\left(n^0\tan{\lambda\over 2}\right)\cr
  \zeta&=2\sinh^{-1}\left[\sqrt{\p0}\sin{\lambda\over 2}\right]\cr
  \psi&=-2\tan^{-1}\left({{n^1}\over{n^2}}\right)={\rm constant}\
     .\cr}
  \eqno(33)$$
It is straightforward to confirm that this solves the geodesic
equation
  $${{d^2x^\mu}\over{d\lambda^2}}+\Gamma^\mu_{\rho\sigma}{{dx^\rho}
  \over{d\lambda}}{{dx^\sigma}\over{d\lambda}}=0 \ ,\eqno(34)$$
for $x^\mu=(\theta,\zeta,\psi)$.  This fact can be seen more
directly by starting with a simple path, such as
$T(\lambda)=\exp(-i\lambda {\J}_0)$, and verifying that this
solves the geodesic equation.  Then by \eq (24) a Lorentz
transformation $\Lambda^\mu{}_\nu$ will take this curve into
another curve of the form $\exp(-i\lambda\phi^\mu
\J_\mu)$, with $\phi^\mu=\Lambda^\mu{}_0$.  Since the action of
SU(1,1) is an isometry, the resulting curve must also be a
geodesic. Finally, the isometry property also ensures that a
curve of the form $T(\lambda)= T_0\exp(-i\lambda n^\mu
\J_\mu)$ ($T_0\in$~SU(1,1)) will be a geodesic through $T_0$.

A simple way to visualize anti-de~Sitter space is in terms of its
Penrose (conformal) diagram \rf{3}.  We define a new coordinate
$\zeta^\prime$ by
  $$\zeta^\prime=4\tan^{-1}\left(e^{\zeta/2}\right)-\pi \
     ,\eqno(35)$$
restricted to the range $0\leq\zeta^\prime<\pi$.  The metric
\eqpar{21} becomes
  $$ds^2={1\over 4 \cos^2 {\zeta' \over 2}} \left(-d\theta^2
     +{d\zeta^\prime}^2 + \sin^2{{\zeta^\prime} \over 2}\
     d\psi^2\right) \ .  \eqno(36)$$
The Penrose diagram is shown in Fig.~5; the angular coordinate
  \def\captione{{\it The conformal diagram of SU(1,1), the double
    cover of SO(2,1).} The group manifold of SU(1,1) with the
    invariant metric is shown.  The $\psi$ direction is
    suppressed; hence, each point away from the $\zeta^\prime=0$
    line represents a circle.  The top and bottom edges,
    $\theta=4\pi$ and $\theta=0$, are identified.  The identity
    element is in the lower left hand corner; we have indicated
    some spacelike and timelike geodesics from this point.
    Elements of SU(1,1) that can be expressed as exponentials of
    generators lie on such geodesics.  The product $T_G$ of two
    such elements $T_B$ and $T_A$ is represented by a curve
    constructed from two consecutive geodesic segments, as shown.
    In this case the product lies in the shaded region, which
    represents elements that cannot be expressed as exponentials
    of generators.  The Gott time machine lies in this region.}%
  \iftwoup
    \figurepageinsert{5}{3in}{4.56in}{0.0in}{\captione}
    {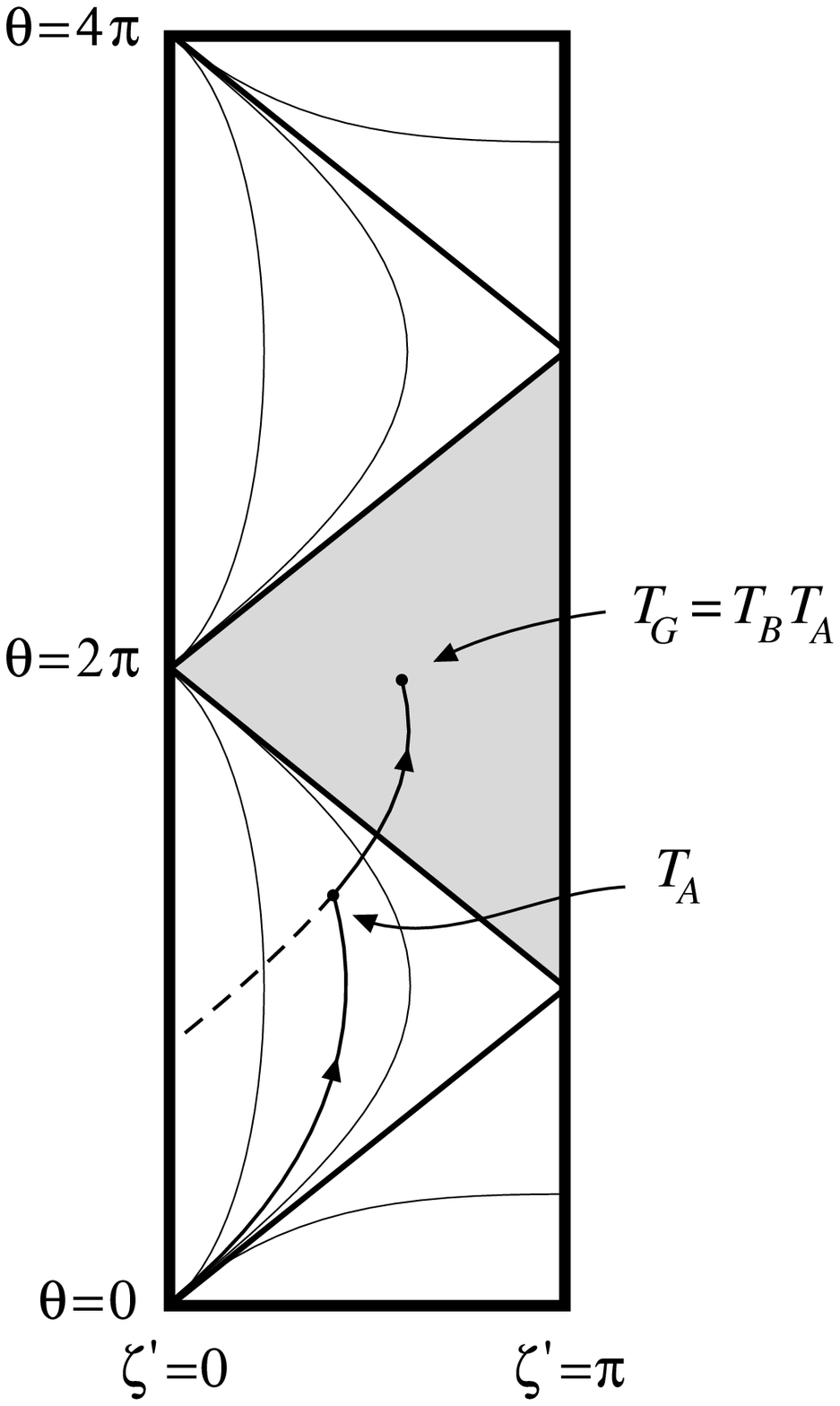 hoffset=-73 voffset=-393 hscale=65 vscale=65}%
  \else
    \figureinsert{5}{3in}{4.56in}{0.0in}{\captione}
    {cfgod5.ps hoffset=-73 voffset=-393 hscale=65 vscale=65}%
  \fi
$\psi$ is suppressed. The light cones at each point are lines
drawn at $45^\circ$.  The right hand side of the rectangle is the
surface $\zeta^\prime=\pi$, which represents spacelike and null
infinity. The lower left corner is the origin, from which we have
drawn typical spacelike and timelike geodesics.  The lower and
upper boundaries are the surfaces $\theta=0$ and $\theta=4\pi$,
which are identified (the topology of SU(1,1) is thus $S^1\times
\IR^2$).  An important feature of this diagram is that timelike
geodesics from the origin refocus at the point ($\theta=2\pi$,
$\zeta^\prime =0$), as can be seen directly from \eq (33) (note
that $\zeta^\prime=0$ is equivalent to $\zeta=0$).  Therefore,
points that are spacelike separated from ($\theta=2\pi$,
$\zeta^\prime =0$) cannot be joined to the origin by a geodesic.
These points correspond to the shaded region of the diagram.

Since every element of SU(1,1) that can be written as the
exponential of a generator lies along a geodesic through the
origin, points in the shaded region correspond to group elements
that cannot be reached by exponentiation.  The element $T_G$
corresponding to the Gott time machine lies in this region.  The
element $-T_G$, on the other hand, can be obtained from $T_G$ by
subtracting $2 \pi$ from $\theta$, so $-T_G$ lies in the region
that is spacelike separated from the origin.  Thus $-T_G$ can be
reached by exponentiation, as was claimed in the previous
section.

The product of two elements $T_B=\exp(-i\phi^\mu_B{\J}_\mu)$ and
$T_A=\exp(-i\phi^\mu_A{\J}_\mu)$ corresponds to traveling in the
direction of $\phi^\mu_A$ along a geodesic to $T_A$, then
traveling along a different geodesic (not through the origin) to
$T_BT_A$. As shown in the diagram, we can easily reach the shaded
region, and hence the Gott time machine, in this manner.  The
parameter space of SO(2,1) can be visualized by cutting the
diagram in half, identifying the surfaces $\theta=0$ and
$\theta=2\pi$.  Then the shaded region of Fig.~5 is mapped onto
the wedge which is covered by spacelike geodesics emanating from
the origin.  This is consistent with our earlier statement that
the Gott time machine holonomy $T_G$, written as an element of
SO(2,1), can be expressed as the exponential of a spacelike
generator, even though the SU(1,1) element for the Gott time
machine cannot be expressed as the exponential of any generator.

Consider a system of particles represented by an element
$T_{tot}$ of SU(1,1).  As in \eq (6), we divide this system into
a subsystem represented by an element $T_{sub}$ and the remaining
$N$ individual particles represented by $T_i$:
  $$T_{tot}= T_N\ldots T_1 T_{sub} \ .\eqno(37)$$
This relation can be represented on the Penrose diagram by a
future-directed nonspacelike curve from $T_{sub}$ to $T_{tot}$,
constructed from geodesic segments representing each of the
$T_i$. It is clear that the periodicity in the $\theta$ direction
allows any two points to be connected in this way --- as far as
SU(1,1) is concerned, any system of particles can contain a
subset with arbitrary energy and momentum.\fn{8}{Note also that
an arbitrary loop, encircling some particles clockwise and others
counterclockwise, will correspond to a sequence of both future-
and past-directed geodesic segments.  However, we will limit our
attention to loops that encircle all particles counterclockwise.}

However, the identification $\theta\leftrightarrow \theta+4\pi k$
obscures an important difference between physically distinct
situations.  To make this difference apparent, we must go to
$\ucg$, the universal cover of SU(1,1) and SO(2,1). In terms of
the Penrose diagram, we no longer identify $\theta=0$ with
$\theta=4\pi$, but instead we extend the picture infinitely far
in the positive and negative $\theta$ direction (Fig.~6).  The
timelike geodesics from the origin will refocus at
  \def\captionf{{\it The universal cover of SU(1,1).}  In the
    universal covering group, the identification of $\theta$ with
    $\theta+4 \pi$ is removed, so that $\theta$ takes values from
    $-\infty$ to $\infty$.  The element $T_G$, representing a
    Gott time machine, does not lie to the past of any point on
    the $\zeta^\prime=0$ line between $\theta=0$ and
    $\theta=2\pi$.  Since all open hypersurfaces with timelike
    total momentum lie on this line (in their rest frame), no
    such hypersurface can contain a Gott time machine.}%
  \ifnum\figurestyle>1
     \vadjust{\bottominsert \figcapt{6:}{\captionf}%
     \endbottominsert}%
  \fi
$\theta=2\pi$, then continue onward, refocusing again at
$\theta=2\pi k$ for every integer $k$.  Therefore the wedges of
points that are spacelike separated from ($\theta=2\pi k$,
$\zeta^\prime=0$) for $k\neq 0$ cannot be reached from any
geodesic through the origin.  All of these points may be said to
correspond to tachyonic momenta, since they map to elements of
SO(2,1) lying on spacelike geodesics from the origin.

     \ifnum\figurestyle>1
     \pageinsert
        \ifnum\figurestyle=3
           \vbox to 0.0in{} \nointerlineskip \centerline{\hbox to
     3.67in{\includegraphics{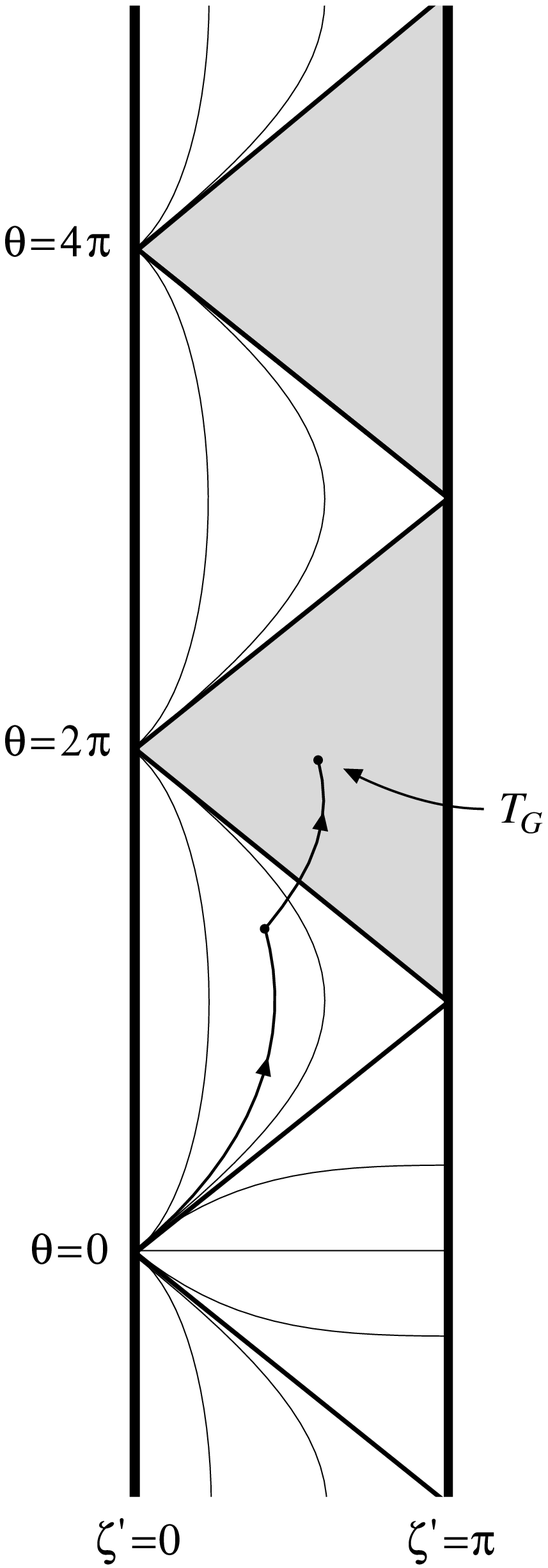}\hfill}} \vfill
        \fi
     \endinsert
     \fi

To describe a multiparticle system using $\ucg$, we express
$T_{tot}$ as a product as in \eq (3), with each $T_i$
representing a single particle.  Any particle in its rest frame
has a deficit angle $\alpha < 2 \pi$, so one can uniquely define
$T_i$ in the universal covering group by using \eq (2),
interpreting $R(\alpha)$ as the element of the universal covering
group described by $(\theta = \alpha, \zeta' = 0)$, where $0 \le
\alpha < 2 \pi$.  Similarly $B(\vecxi)$ can be chosen to lie in
the sector of the universal covering group that is spacelike
separated from the identity. (In this case, however, any other
choice would be equivalent.  The ambiguity consists of any number
of factors of the group element corresponding to a rotation by $2
\pi$, and this element commutes with all other elements.
Therefore, in \eq (2), the ambiguous factor would cancel between
$B$ and $B^{-1}$.)

We will continue to use the word ``holonomy'' to refer to the
group element $T_{tot} \in \ucg$, although there is no type of
physical particle that we could parallel transport around a
closed loop which would distinguish between elements of $\ucg$
separated by a $4\pi$ rotation.  Unlike a true holonomy, however,
the $\ucg$ element is not uniquely defined by the loop $C$ alone:
it is also necessary to specify, up to continuous deformation, a
disk $D_C$ of which $C$ is the boundary.  This disk allows a
unique specification of which particles are inside the loop, so
that \eq (3) can be written.  (If the loop $C$ were drawn on an
$S^2$ surface, for example, there would be two inequivalent areas
that it would bound.  The two resulting elements of $\ucg$ would
differ by a $4 \pi$ rotation, so the specification of $D_C$ is
needed to resolve the ambiguity.)

The definition of $T_{tot} \in \ucg$ given above is conveniently
explicit, but we must show that the result is independent of the
arbitrary paths $P_i$ that were introduced to write $T_{tot}$ as
a product of single-particle group elements $T_i$, as was done in
\eq (3).  For this purpose, we introduce an alternative
definition.  Recall that an element of the covering group $\tilde
G$ of any group $G$ is uniquely defined by specifying an element
$g \in G$, and by specifying, up to continuous deformation, a
path in $G$ connecting the identity element to $g$.  To make use
of this fact, imagine that the mass of each particle is rendered
nonsingular by smearing it over a small region.  To define the
$\ucg$ holonomy of a loop $C$, start with a trivial loop in $D_C$
(i.e., one which encircles no particles), and deform it through
$D_C$ into the loop $C$. At each stage of the deformation there
is an unambiguously defined holonomy element of SO(2,1), and
therefore the deformation produces a continuous path in SO(2,1)
from the identity to the holonomy group element for the loop $C$.
Although the precise path in SO(2,1) will depend on the loop
deformation that is chosen arbitrarily, the specification of
$D_C$ guarantees that the path will be determined up to
continuous deformation, precisely what is needed to define an
element of the covering group $\ucg$.  Finally, this definition
can be shown equivalent to the one given two paragraphs above by
considering in particular the deformation of the loop which
starts as a trivial loop, and expands to cover each of the
particles one at a time, in the same order as the factors
appearing in \eq (3).

Since the element $T_{tot} \in \ucg$ depends only on $C$ and
$D_C$, relationships derived by the continuous deformation of
loops, such as the conservation law of \eq (4) or the isolation
of a subsystem described by \eq (37), are valid equations in
$\ucg$.

We can now describe the key step in our argument that a Gott time
machine cannot be created from initial conditions specified on an
open spacelike hypersurface with timelike total momentum.
Consider an arbitrary subset (possibly the entire set) of
particles crossing a spacelike hypersurface $S$, with a momentum
characterized by a holonomy element $T_{sub}$ of $\ucg$. Then
\eqpar{37} implies that the holonomy of all particles crossing
$S$ can be written as
  $$T_{tot}=T_N\ldots T_1 T_{sub}\ ,\eqno(38)$$
where the $T_i$ represent the individual particles comprising the
rest of the system.  The right hand side of this expression must
correspond to a point in the future light cone of $T_{sub}$,
since each $T_i$ can be represented by a segment of a
future-directed timelike or null geodesic.  For an open
hypersurface with timelike total momentum, however, $T_{tot}$ in
the rest frame corresponds to a rotation by the total deficit
angle, which we will show must be less than or equal to $2\pi$.
Thus, $T_{tot}$ must lie on the $\zeta'=0$ axis, between
$\theta=0$ and $\theta=2\pi$. However, the holonomy element of a
Gott time machine, $T_G$, lies in the shaded region that is
spacelike separated from the point $(\theta = 2 \pi, \zeta'=0)$,
as shown in Fig.~6.  Since $T_{sub}$ must lie to the past of
$T_{tot}$, $T_{sub}$ cannot represent the momentum of a Gott time
machine.  Thus, no subset of the particles crossing $S$ can
comprise a Gott time machine.

To complete the argument, we must investigate the circumstances
under which one can find a spacelike hypersurface $S$, as was
used in the previous argument.  The complication is the
possibility of CTC's, which cannot be discounted {\it a priori},
as we are trying to prove that they cannot be created.  In the
presence of CTC's a spacetime will not necessarily admit a
foliation into spacelike hypersurfaces, so it may not be possible
to choose a holonomy loop so that all of the particles contribute
positively.  In addition, we must prove that the total deficit
angle of an open hypersurface must be less than or equal to $2
\pi$.  These steps are completed in the next section.

\head{III. RESTRICTIONS ON TIME MACHINE CONSTRUCTION}

In this section we consider an attempt to build a Gott time
machine in a universe that is open, in the sense that it
contains\fn{9}{In the absence of CTC's, it is reasonable to
define open universes as those which may be foliated by spacelike
surfaces with $\IR^2$ topology.  Such a foliation may not be
possible when CTC's are present, however, even in spacetimes
which we would intuitively classify as ``open.'' Hence, we will
take the existence of a single such surface as the defining
characteristic of an open universe.} an edgeless spacelike
surface $S$ with the topology of $\IR^2$. We will further limit
our attention to the case where $S$ is acausal (no timelike or
null curve intersects $S$ more than once); in that case $S$ is
called a partial Cauchy surface.  Then we can speak of the total
momentum of the particles passing through $S$, defined by the
element $T_S$ of $\ucg$ computed from the holonomy of a
non-self-intersecting loop in $S$ at infinity, enclosing all of
the particles.  When all of the particles in the universe
intersect $S$, we may think of $T_S$ as defining the total
momentum of the universe; we will usually consider this
situation, although it is not strictly necessary to our argument.
In the case where $T_S$ is timelike, we will show that a group of
particles with the momentum of a Gott time machine can evolve
from the data specified on $S$ only if the rest frame deficit
angle is greater than $2\pi$.  We then show that this deficit
angle must be less than or equal to $2\pi$ if $S$ has the
topology of $\IR^2$, {\it i.e.}, if the (2+1)-dimensional
universe is open.

We begin with some basic definitions; more complete discussions
can be found in Refs.~\refbrack{2}, \refbrack{3}, \refbrack{5},
and \refbrack{21}.  Consider a spacetime containing a partial
Cauchy surface $S$.  The future domain of dependence $D^+(S)$ is
the set of all points $p$ such that every past-directed
inextendible timelike or null curve through $p$ intersects $S$.
Thus, initial data specified on $S$ suffice to determine the
evolution throughout $D^+(S)$. The past domain of dependence, and
analogous terminology, is defined in the obvious way and denoted
by replacing the plus sign by a minus sign. The full domain of
dependence, $D(S)$, is defined as $D^+(S)\cup D^-(S)$. The future
boundary of $D^+(S)$, past which information specified on $S$ is
no longer sufficient to determine the evolution, is a null
surface called the future Cauchy horizon $H^+(S)$.  There are
various circumstances under which a Cauchy horizon may arise,
including the creation of CTC's.  Any point $p$ that lies on a
CTC in the future of $S$, or in the future of any such CTC, is
not contained in the domain of dependence $D^+(S)$, since there
will exist a past-directed inextendible timelike curve through
$p$ which wraps forever around a CTC without ever intersecting
$S$.

We also define a partial ordering in the group $\ucg$: we say
that $T_i < T_j$ if and only if $T_i$ lies in the past light cone
of $T_j$, in the anti-de Sitter metric of \eq (21).  The light
cones used in this definition are easily visualized by using the
conformal diagram of \fig 6.

We will show that if the momentum associated with $S$ is
timelike, then no Gott time machine can evolve from the particles
passing through $S$.  That is, no group of particles in the
future domain of dependence $D^+(S)$, or on the Cauchy horizon
$H^+(S)$, can have the holonomy of a Gott time machine.  Thus,
either the Cauchy horizon does not exist, or it arises due to an
effect other than the creation of a Gott time machine.

First let $S'$ be any Cauchy surface for $D(S)$, and let $C$ be a
non-self-intersecting loop on $S'$, which defines a holonomy
$T_C$. We may decompose $T_{S'}$ into $T_C$ times the positive
contributions of other particles passing through $S'$, as in
\eqpar{37}.  Hence $T_C$ must be less than or equal to $T_{S'}$,
in the ordering defined above.  But since $S'$ is a Cauchy
surface, the particles that cross $S'$ are exactly those that
cross $S$.  The loop defining $T_{S'}$ can therefore be deformed
into the loop defining $T_S$ without crossing any particle
worldlines. This deformation can be carried out even if the
particles undergo merges or decays, so in all cases $T_{S'} =
T_S$.  Thus, $T_C
\le T_S$.

Now consider the general case where $S'$ is any connected
edgeless spacelike surface in $D(S)$ and $C$ is again a
non-self-intersecting loop on $S'$.  We will show that $T_C \le
T_S$ by showing that $T_C$ can be embedded in a Cauchy surface
$S''$, reducing this situation to the case of the previous
paragraph.

We begin by using the method of Geroch \rf{22} to
define\fn{10}{In \Ref{22}, Geroch restricts his construction to
the interior of $D(S)$.  However, with the present definition of
$D^+(S)$ (which follows Refs.~\refbrack{3} and \refbrack{21} and
differs slightly from that of Geroch) the condition that $S$ be
edgeless implies that $D(S)$ is open.  The construction,
therefore, applies to all of $D(S)$.} a time coordinate $\lambda$
on $D(S)$.  Choose a measure on $D(S)$ so that the total volume
of $D(S)$ is 1, and let $V^+(p)$ denote the volume of the future
of a point $p$ in this measure.  Similarly, let $V^-(p)$ denote
the volume of the past, and let $\lambda \equiv V^-/V^+$.
Clearly $\lambda$ increases on any timelike curve. Geroch shows
that $\lambda$ is continuous and that it takes all values from
$-\infty$ to $\infty$ on any inextendible timelike curve.

We now construct a coordinate system on $D(S)$ as follows: let
$(x,y)$ be coordinates for $S$, and then let $(\lambda_0, x, y)$
denote the point where $\lambda=\lambda_0$ on the integral curve
of $\nabla\lambda$ through $(x,y)$.  Since $S$ has the topology
of $\IR^2$, $D(S)$ has the topology of $\IR^3$, and in particular
is simply connected.

We now show that since $S'$ is spacelike, edgeless, and
connected, it divides $D(S)$ into two regions: the past and the
future of $S'$.  For any point $p\in D(S)$, we let $\Gamma
\subset D(S)$ be a curve from $p$ that intersects $S'$ at least
once, and we let $p'$ denote the first such intersection point.
The point $p$ will be said to lie to the future or the past of
$S'$ according to whether $\Gamma$ approaches $p'$ from the
future or the past side of $S'$. This is well-defined, because if
two curves $\Gamma$ and $\Gamma'$ from $p$ approach $S'$ from
different sides, we would be able to join $\Gamma$ and $\Gamma'$
by a curve through $S'$ to obtain a closed curve.  An arbitrarily
small distortion would then produce a curve that crosses $S'$
exactly once.  But such a curve is impossible: the topology of
$D(S)$ implies that the curve would be contractible, but the
edgelessness of $S'$ implies that the number of crossings can
only change by an even number.\fn{11}{See, for example, p.~204 of
\Ref{3}.}  One can similarly show that $S'$ is achronal, since
any nonspacelike curve connecting two points on $S'$ could also
be used to construct a closed curve that crosses $S'$ exactly
once.

  \def\captiong{{\it The construction of the Cauchy surface
    $S''$.}  Given an edgeless spacelike surface $S'$ in $D(S)$
    and a closed region $D_C \subset S'$, the procedure described
    in the text serves to define another surface $S''$ such that
    $D_C \subset S''$, and $S''$ is a Cauchy surface for $D(S)$.
    In this figure, $S_{\rm min}$ and $S_{\rm max}$ are surfaces
    defined by $\lambda=\lambda_{\rm min}$ and $\lambda =
    \lambda_{\rm max}$, respectively, where $\lambda_{\rm min}$
    and $\lambda_{\rm max}$ are the minimum and maximum values of
    the function $\lambda$ on the region $D_C$.}

The loop $C$ and the interior of $C$ in $S'$ together comprise a
compact region $D_C$.  Within $D_C$, $\lambda$ assumes a maximum
value $\lambda_{\rm max}$ and a minimum value $\lambda_{\rm
min}$.  The desired surface $S''$ can then be defined by the
relation $\lambda = f(x,y)$, where the function $f(x,y)$ is
defined by the following procedure, illustrated in \fig 7. If for
the given $x$ and $y$ there is a point $(\lambda, x, y) \in S'$
with $\lambda_{\rm min}\le\lambda\le\lambda_{\rm max}$, then let
$f(x,y) = \lambda$.  If not, then the points $(\lambda_{\rm min},
x, y)$ and $(\lambda_{\rm max}, x, y)$ are both on the same side
of $S'$. If they are in the past, let $f(x,y)=\lambda_{\rm max}$;
if they are in the future let $f(x,y)=\lambda_{\rm min}$.  Note
that $f(x,y)$ is continuous, and is bounded by $\lambda_{\rm min}
\le f(x,y) \le \lambda_{\rm max}$.
  \figureinsert{7}{4.19in}{1.87in}{0.0in}{\captiong}
    {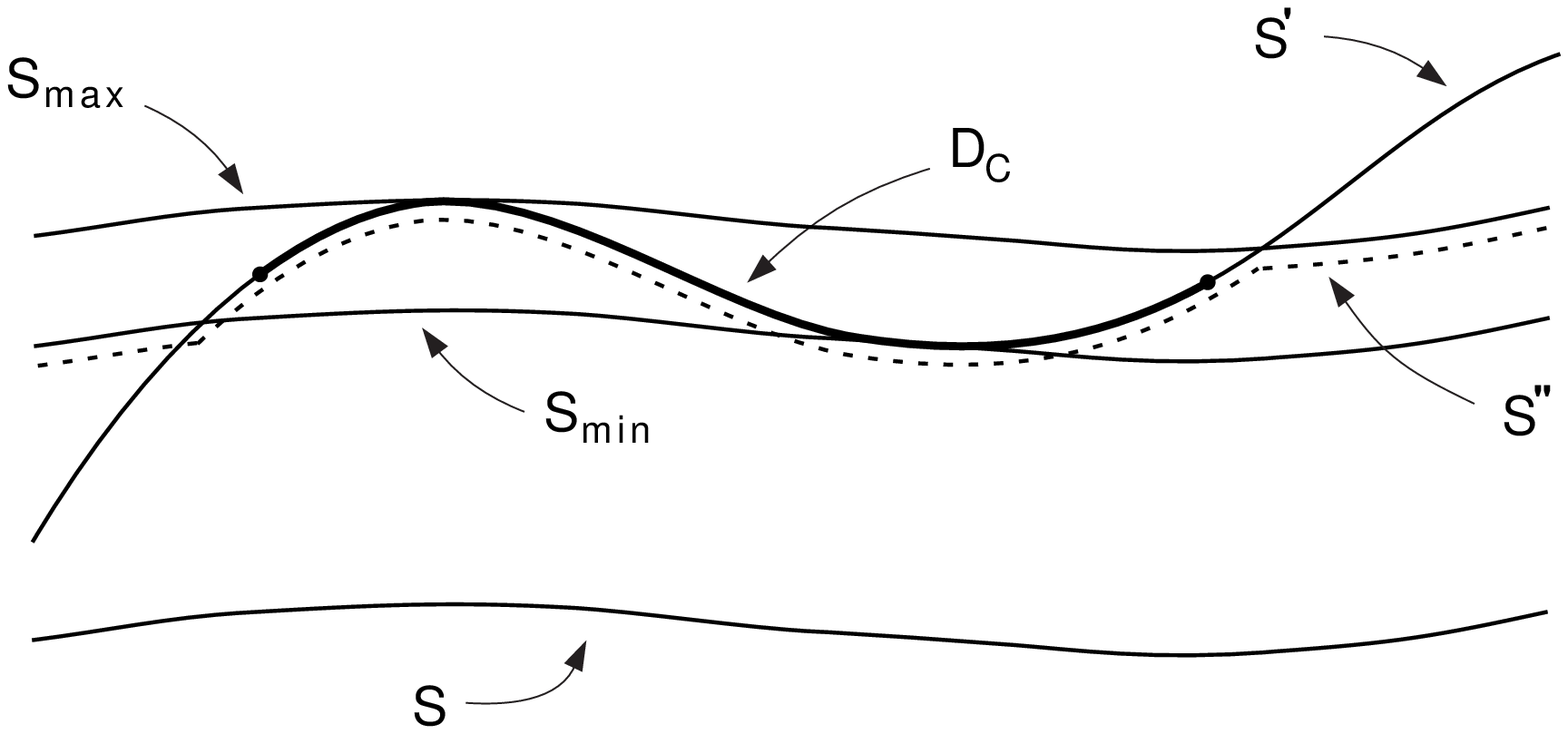 hoffset=-10 voffset=-269 hscale=57 vscale=57}

To complete the argument, we must show that $S''$ is a Cauchy
surface for $D(S)$, {\it i.e.}, that any inextendible timelike
curve $\gamma$ intersects $S''$ exactly once.  For any point
$(\lambda, x, y)$, let $\tilde \lambda = \lambda - f(x,y)$. The
surface $S''$ is then described by $\tilde \lambda = 0$, with
$\tilde \lambda$ positive in the future and negative in the past.
Since $\lambda$ takes all values from $- \infty$ to $\infty$
along the curve $\gamma$, the same must be true for $\tilde
\lambda$, and hence $S''$ must be intersected at least once.
Following $\gamma$ toward the future, we see that at the first
crossing of $S''$, $\tilde \lambda$ changes from a negative value
to a positive one.  But then there can be no further crossings,
since at the next crossing $\tilde \lambda$ would have to change
from positive to negative, which implies that $\gamma$ would
cross the spacelike surface $S''$ toward the past. Thus there is
exactly one intersection of $\gamma$ and $S''$, so $S''$ is a
Cauchy surface for $D(S)$.  Since $C$ lies on $S''$, we find by
the earlier argument that $T_C \le T_S'' = T_S$.

  \def\captionh{{\it Holonomy of a loop on the Cauchy horizon.}
    The loop $C$ is joined by a timelike cylinder $\Delta$ to a
    loop $C^\prime$ embedded in $S^\prime$, a spacelike surface
    in the domain of dependence of $S$.  No particles pass
    through $\Delta$, so the holonomy of $C$ is equal to that of
    $C^\prime$, which arises from the contributions of particles
    passing through the shaded region, the interior of $C^\prime$
    in $S^\prime$.}

An equivalent result holds for particles passing through the
Cauchy horizon.\fn{12}{$H^+(S)$ is null, rather than spacelike,
but the definition of the holonomy of a system of particles can
be trivially extended to this case, as long as no particle
worldline lies on $H^+(S)$.  If such a particle worldline did
exist, the holonomy would not be defined, since the holonomy loop
would intersect it.} To see this, consider a
non-self-intersecting loop $C$ on $H^+(S)$. We will deform $C$
into the interior of $D(S)$, where the previous result can be
used.  Let $V$ be any continuous timelike vector field defined on
$D^+(S) \cup H^+(S)$.  From each point of $C$, follow the
integral curve of $V$ backward through $D^+(S)$ until it reaches
a surface $S'$ defined by $\lambda=\lambda'$, where $\lambda'$ is
a constant. Call the resulting loop $C'$, and let $\Delta$ be the
cylinder that connects $C$ to $C'$, as shown in \fig 8.
  \figureinsert{8}{4.32in}{2.64in}{0.0in}{\captionh}
    {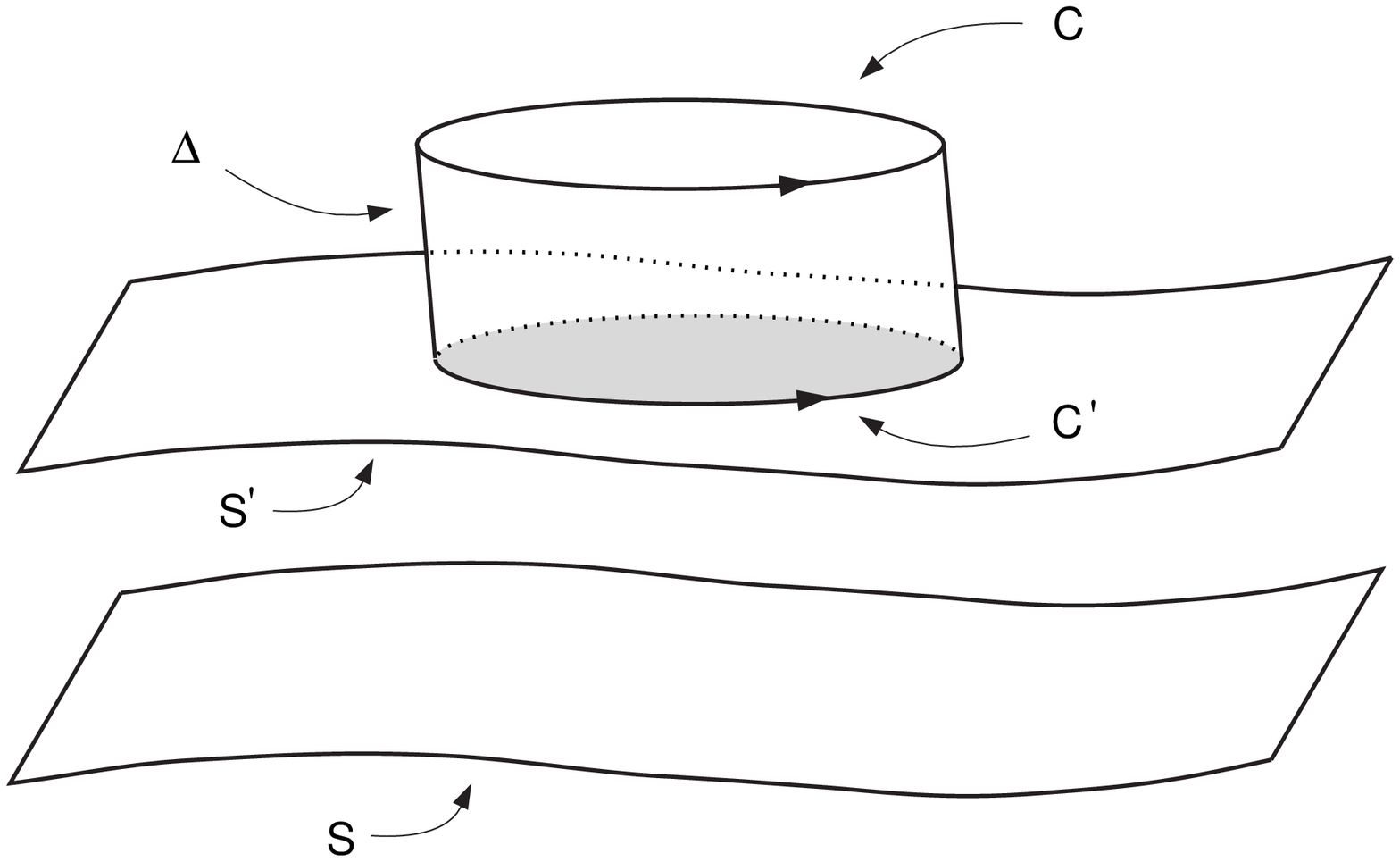 hoffset=-8 voffset=-262 hscale=57 vscale=57}%
Choose $\lambda'$ large enough ({\it i.e.}, $S'$ late enough) so
that no particle worldline intersects $\Delta$.  The holonomy
$T_C$ is then equal to $T_{C'}$, which is less than or equal to
$T_S$ by the argument above.

We will now show that any spacetime containing a complete
spacelike surface with the topology of $\IR^2$ and a timelike
total momentum must have a rest frame deficit angle less than or
equal to $2\pi$.  We first consider the case of an arbitrary
2-dimensional surface that is flat everywhere except for the
interior of a loop $L$; only the intrinsic properties of the
surface will concern us in this paragraph. Let $\Sigma$ be the
interior of $L$, and $K$ be the scalar curvature of $\Sigma$. The
Gauss-Bonnet theorem (see, e.g., \rf{23}) tells us that, for
surfaces with the topology of $\IR^2$,
  $$\int_\Sigma K\ dA=2\pi-\int_L{{d\phi}\over{ds}}\ ds\ .\eqno(39)$$
Here, $\phi$ is the angle between the tangent vector of $L$ and
an arbitrary vector parallel transported along $L$ in the
2-dimensional surface.  We call the right hand side of this
equation the turning deficit angle $\theta_{\rm turn}$, and we
will show below that it is equal to the holonomy deficit angle
$\theta$, defined earlier. As the surface is flat outside the
loop (by hypothesis), it follows that the left hand side of \eq
(39) gives the integrated curvature over the entire surface.  We
can then invoke a theorem of Cohn-Vossen \rf{13}, which states
that the integrated curvature of a geodesically complete surface
with the topology of $\IR^2$ is less than or equal to $2\pi$,
from which it follows that
  $$\theta_{\rm turn} \equiv 2\pi-\int_L{{d\phi}\over{ds}}\ ds
     \leq 2 \pi\ .\eqno(40)$$
Thus, a two-dimensional surface that is flat outside of a loop
$L$ and has $\IR^2$ topology necessarily has a turning deficit
angle less than or equal to $2\pi$.

Our task, then, is to find such a surface in our spacetime.  The
surface $S$ whose existence we have hypothesized is not
necessarily sufficient --- even though the portion of $S$ that
lies outside the loop $L$ is locally embedded in Minkowski space,
there is no guarantee that it is intrinsically flat, and it is
therefore hard to apply the above reasoning.  However, it is
possible to define a flat metric on this portion of $S$, by the
following procedure.  Let $\hat t$ denote the future-directed
unit vector left invariant by the holonomy around $L$.  If the
holonomy is a rotation by a multiple of $2 \pi$, then choose
$\hat t$ to be any future-directed timelike unit vector.  In
either case, $\hat t$ can be consistently parallel transported
throughout the portion of $S$ exterior to $L$.  In a
(three-dimensional) neighborhood of any point on $S$, we can
construct a Minkowskian coordinate system $(t,x,y)$ with metric
$ds^2=-dt^2+dx^2+dy^2$, such that the direction of the $t$-axis
coincides with $\hat t$. Since $S$ is spacelike, we may use
$(x,y)$ as coordinates on $S$ locally, and then define the metric
on the region covered by these coordinates to be $dx^2+dy^2$.
Note that this need not agree with the induced metric from the
spacetime, as $S$ may be curved in the latter metric.  Note also
that the flat metric is uniquely defined: the local Minkowskian
coordinate system is specified up to translations or rotations in
the $x$-$y$ plane, under which the metric $dx^2+dy^2$ is
invariant.  We thus have a well-defined flat metric on $S$,
outside the loop $L$.

Now we must show that the turning deficit angle $\theta_{\rm
turn}$ is equal to the holonomy deficit angle $\theta_S$, defined
by the rotation angle of $T_S \in \ucg$ in its rest frame.  For
this purpose we will use the alternative definition discussed in
\sec II for defining the holonomy $T_S$ in the universal covering
group.  By this definition, the ambiguity of rotations by $2 \pi$
that occurs in SO(2,1) is resolved by continuous deformation of
the loop.  We introduce a 1-parameter class of loops $L_\lambda$,
each with the same base point $Q$, where $L_0$ is a trivial loop
that encircles no particles and $L_1 = L$.  As in \sec II we
imagine that the mass of each particle is smeared over a small
region, so the holonomy changes continuously as the loop is
varied.  Let $T_\lambda$ denote the holonomy of $L_\lambda$, and
note that it is uniquely defined in $\ucg$ by requiring that it
be continuous in $\lambda$, with $T_0 \equiv I$ (the identity
element).  For definiteness, we describe the loop $L_\lambda$ in
parameterized form as $L_\lambda(s)$, where $0 \le s \le 1$.  Let
$V_\lambda(s)$ denote the tangent vector of $L_\lambda$ at $s$,
mapped to the tangent space at $Q$ by parallel transporting
backward (i.e., clockwise) along $L_\lambda$.  Thus,
$V_\lambda(s)$ traces out a continuous curve in the tangent space
at $Q$.  The tangent vector must return to its original value at
$s=1$, but $V_\lambda(1)$ is defined so that its value is
modified by parallel transport clockwise around the loop
$L_\lambda$.  Inverting this transformation, we have
  $$T_\lambda \, V_\lambda(1) = V_\lambda(0)\ .\eqno(41)$$

\eq (41) allows us to define a continuous loop in the tangent
space at $Q$ by noting that $\ucg$ is simply connected, so a
curve $g_\lambda(s)$ in $\ucg$ satisfying $g_\lambda(1)=I$ and
$g_\lambda(2)=T_\lambda$ can be constructed uniquely, up to
continuous deformation.  Then define a closed curve
$\vbar_\lambda(s)$ in the tangent space at $Q$ by concatenating
$V_\lambda(s)$ and $g_\lambda(s) V_\lambda(1)$:
  $$\vbar_\lambda(s) \equiv \cases{V_\lambda(s) &if $0 \le
     s \le 1$\cr g_\lambda(s) V_\lambda(1) &if
     $1 \le s \le 2 $\ .\cr}\eqno(42)$$
$\vbar_\lambda(s)$ is confined to the spacelike part of the
tangent space, which is not simply connected (note that the zero
vector is not spacelike).  $\vbar_0(s)$ makes one
counterclockwise loop around the origin as $s$ is varied from 0
to 2, so by continuous deformation the same statement must hold
for hold for all $\lambda$, and in particular for $\lambda=1$.

Now we must connect the behavior of the 3D tangent vector $V$
(defined in the (2+1)-dimensional tangent space at $Q$) to the 2D
tangent vector in the surface $S$, which was used to define the
turning angle.  We use the same local Minkowskian coordinate
system $(t,x,y)$ that was used to construct the flat metric, and
again we take $x$ and $y$ to locally define coordinates on $S$.
The components of the 2D tangent vector are then equal to the $x$
and $y$ components of the 3D tangent vector, and in both the 2D
and 3D spaces these components are unchanged by parallel
transport.  For any infinitesimal segment of loop, the turning
angle $d \phi$ of the 2D tangent vector, as calculated in the
flat metric, is equal to the change in the azimuthal angle of the
3D tangent vector (i.e., the angular change of the projection of
the 3D tangent vector into the $x$-$y$ plane).  Thus, the
integrated change in the azimuthal angle of the tangent vector
$\vbar_1(s)$ for $0 \le s \le 1$ is equal to the total turning
angle,
  $$\int_L{{d\phi}\over{ds}}\ ds \ .$$
Since $g_1(1)=I$ and $g_1(2) = T_S$, the integrated change in the
azimuthal angle of $\vbar_1(s)$ for $1 \le s \le 2$ is equal to
the holonomy deficit angle $\theta_S$.  Since $\vbar_\lambda$ is
known to make one counterclockwise loop around the origin, we
have
  $$\int_L{{d\phi}\over{ds}}\ ds + \theta_S = 2 \pi \ ,\eqno(43)$$
which shows that $\theta_{\rm turn} = \theta_S$.

In the case at hand, this suffices to show that the holonomy
$T_S$ must lie on the conformal diagram of $\ucg$ in the triangle
to the future of the identity and the past of the region
describing Gott time machines.  Since we have argued that the
holonomy of any group of particles in the domain of dependence of
$S$, or on the Cauchy horizon (if it exists), must have a
holonomy less than or equal to $T_S$, it is not possible for
these particles to comprise a Gott time machine.  We have
therefore established that Gott time machines cannot be created
in open universes with timelike total momentum.

\head{IV. CONCLUSIONS}

The role played by closed timelike curves is an important issue
in classical general relativity, and may be important in an
ultimate quantum version of the theory.  The general theorems of
Tipler and Hawking are strong statements about the difficulty of
creating CTC's, but incomplete in that they do not specify what
will go wrong with any particular attempt at time machine
construction.  In this paper we have studied some specific
obstacles to the creation of time machines of the type discovered
by Gott \rf{8}, using the considerable simplification afforded by
working in the toy model of (2+1)-dimensional gravity.  These
obstacles are most easily understood by considering the
anti-de~Sitter geometry of the 3-dimensional Lorentz group, in
which we find that Gott time machines cannot lie to the past of
collections of particles with timelike momentum and deficit angle
less than $2\pi$. We then use this fact to show that a Gott time
machine cannot be created in an open universe with a timelike
total momentum, essentially because there can never be enough
energy in an open universe to achieve the Gott condition.

This result is situated within an ongoing discourse concerning
the appearance and significance of CTC's in general relativity,
including discussions of whether physics can be consistent in the
presence of CTC's \rf{24}.  Considerable effort has recently been
invested in understanding the creation of CTC's in
(3+1)-dimensional spacetimes with traversable wormholes \rf{25}.
Such spacetimes seem to easily develop CTC's, but the maintenance
of a traversable wormhole requires violation of the weak energy
condition (WEC).  While quantum field theory in curved spacetime
can allow WEC violation, there is evidence that quantum
fluctuations serve to destabilize the would-be time machine,
preventing the appearance of CTC's \rf{4,26}.  This issue has led
to several investigations of the behavior of quantum fields on
background spacetimes with CTC's \rf{27}.  These studies, which
ask whether CTC creation is possible when the WEC is relaxed, are
complementary to the one presented in this paper, which examines
the nature of obstacles to CTC creation when the WEC is enforced.

It is unclear, however, what the implications of our result are
for CTC creation in the real (3+1)-dimensional world.  We have
seen that in (2+1) dimensions, where the unique property that
spacetime is flat in vacuum precludes the possibilities of black
hole creation and energy loss through gravitational radiation,
this same property leads to a restriction on the total energy of
an open universe with timelike total momentum, which in turn
presents insuperable obstacles to time machine creation.
However, the notion of a timelike momentum is rather intimately
connected with the nature of the (2+1)-dimensional theory, so it
seems inevitable that the Tipler-Hawking theorems in (3+1)
dimensions must be enforced by other means.  Thus, a general
understanding of the status of time machine creation remains
elusive.

The issue of closed timelike curves in (2+1) dimensions,
moreover, has not been completely resolved; while we have pointed
out in Section~III that the Tipler-Hawking theorems apply in this
context, there remains the possibility of time machines (distinct
from the type proposed by Gott) with non-compactly generated
Cauchy horizons. Waelbroeck \rf{28} has shown that a two-particle
system with timelike momentum does not support CTC's, and Kabat
\rf{29} has presented arguments suggesting that this result is
more general.  Menotti and Seminara \rf{30} have discovered
restrictions on the existence of time machines in stationary and
axially symmetric spacetimes. A comprehensive proof (or
counterexample) is worth searching for.

Meanwhile, the interpretation of holonomies in terms of the
anti-de~Sitter geometry of SO(2,1) sheds light on the
``tachyonic'' nature of the Gott two-particle system.  While the
energy-momentum vector of such a pair is properly described as
spacelike, this fact does not render such a pair unphysical, as
the energy-momentum vector does not tell the entire story. In the
universal covering group of SO(2,1), the region containing Gott
pairs is disjoint from that containing tachyons.  The obstacle to
creating a Gott time machine is not the tachyonic momentum as
such, but the absence of sufficient energy in an open universe.
This is seen most clearly by considering the case of closed
universes.  In our earlier paper \rf{14} we argued that it is
impossible to produce two particles satisfying the Gott condition
by the decay of slowly-moving parent particles unless the total
rest frame deficit angle exceeds $2\pi$.  In a closed universe,
where the total deficit angle is $4\pi$, this does not constitute
an obstacle.  It is easy to construct a closed universe containing
two particles, each with deficit angle between $\pi$ and $2\pi$,
and a number of less massive spectator particles which bring the
total deficit angle to $4\pi$.  Using the description of decays
given in \Ref{14}, we have found that the two massive particles
can decay in such a way that each emits an offspring at
sufficiently high velocity that the total momentum of the two
fast-moving particles is tachyonic.\fn{13}{Our construction is
described by 't Hooft in \Ref{17}.}  Thus, in a closed
(2+1)-dimensional universe it is possible to ``build a tachyon.''
However, as we mentioned in the introduction, 't~Hooft \rf{17}
has shown that the size of the universe begins to shrink after
the decays, leading to a crunch (zero volume) before any CTC's
can arise.  There is thus a sense in which general relativity is
flexible enough to permit tachyons, but works very hard to
prevent time travel.

\goodbreak
\bigskip \bigskip
\centerline{\bigbf ACKNOWLEDGEMENTS}
\nobreak
\medskip

We thank Bruce Allen, Malcolm Anderson, Raoul Bott, Robert Finn,
John Friedmann, Alex Lyons, Robert MacPherson, Don Page, Ted
Pyne, Jonathan Simon, Isadore Singer, Gerard 't~Hooft, and David
Vogan for very helpful conversations. The work of S.M.C. was
supported in part by funds provided by the U.S. National
Aeronautics and Space Administration (NASA) under contracts
NAGW-931 and NGT-50850 and the National Science Foundation under
grant PHY/9206867; the work of S.M.C., E.F. and A.H.G. was
supported in part by the U.S. Department of Energy (D.O.E.) under
contract \#DE-AC02-76ER03069; and the work of K.D.O. was
supported in part by the Texas National Research Laboratory
Commission under grant \#RGFY93-278C.

\goodbreak
\bigskip \bigskip
\centerline{\bigbf APPENDIX A: THE TIPLER/HAWKING THEOREMS}
\nobreak
\medskip

In this Appendix we review the theorems of Tipler \rf{5} and
Hawking \rf{4}, who make rigorous the notion of ``building a time
machine in a local region of spacetime,'' and then show that such
construction is impossible using only normal matter, in the
absence of singularities.  We also explain how the theorems may
be extended to (2+1) dimensions.

The future Cauchy horizon $H^+(S)$, defined in Sec.~III as the
boundary of the future domain of dependence $D^+(S)$, is a null
surface which can be thought of as the union of null geodesic
segments known as ``generators.'' The generators of $H^+(S)$ have
no past endpoints, and they never intersect in the past.  They
may have future endpoints if the generators intersect; the set of
such endpoints forms a set of measure zero.  The notion of
``creating a time machine in a local region'' can be defined
precisely by considering Cauchy horizons for which every
generator, when followed into the past, enters a compact region
of spacetime $B$ and remains there.  (Note that a set which would
otherwise be compact can be rendered non-compact by the
appearance of a curvature singularity.) Hawking \rf{4} refers to
such horizons as ``compactly generated.'' These are the types of
time machines that could, in principle, be constructed by an
advanced civilization.  Since the generators can have no past
endpoints, each generator entering the region $B$ must wind round
and round within $B$ \rf{4}.

The situation is thus as we have portrayed in \fig 9. A compact
set $B$ lies in the future of a partial
  \def\captioni{{\it Time machine creation in a local region.} A
    partial Cauchy surface $S$ is pictured, in the future of
    which CTC's evolve.  The Cauchy horizon, labelled $H^+(S)$,
    emerges from a compact set $B$.}%
  \figureinsert{9}{4.18in}{3.08in}{0.0in}{\captioni}
    {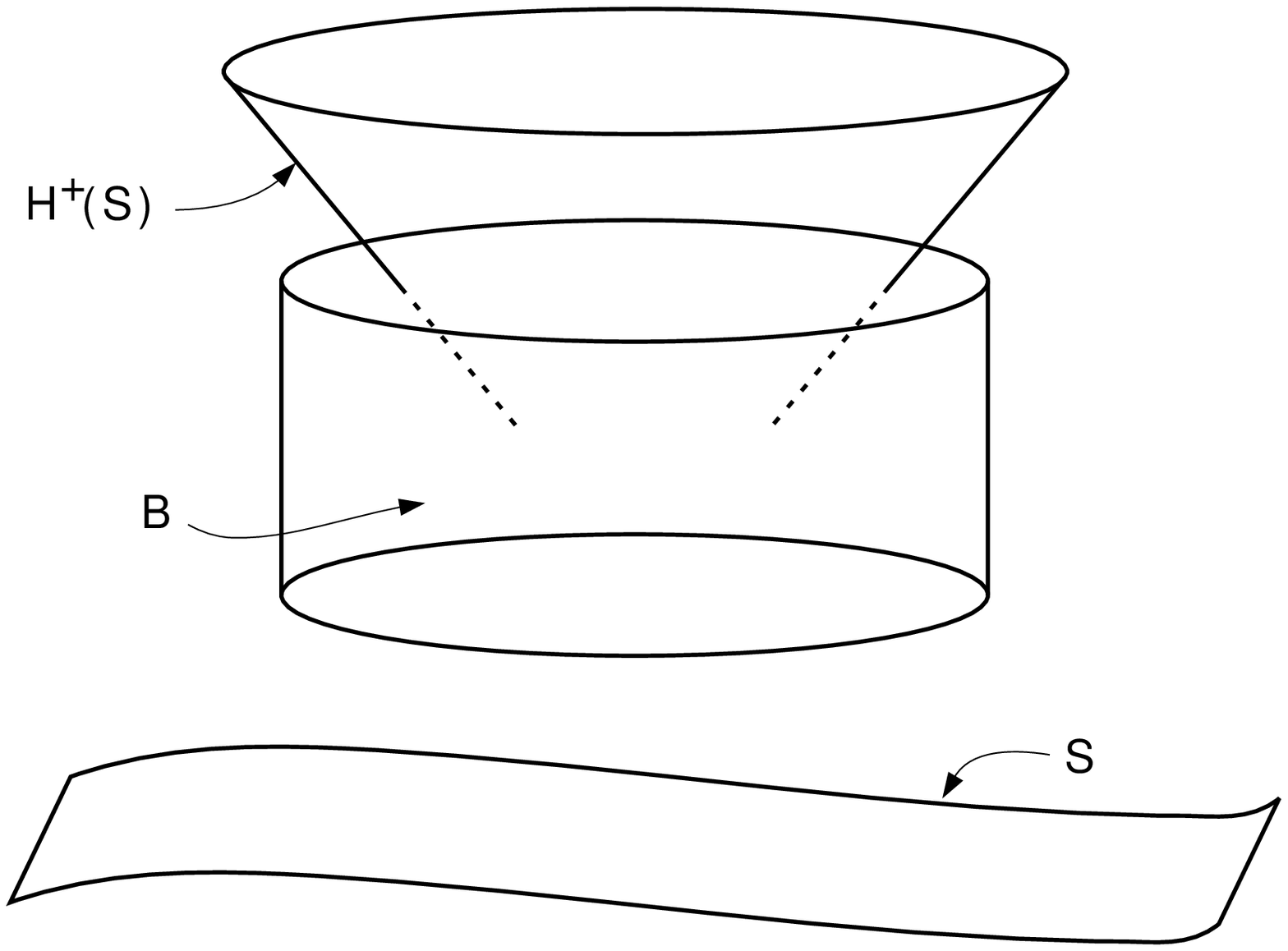 hoffset=-17 voffset=-340 hscale=62 vscale=62}%
Cauchy surface $S$.  The Cauchy horizon $H^+(S)$ emerges from a
compact set $B$, which may be thought of as the place where CTC's
are created.  Tipler and Hawking essentially prove that this
picture will never describe a spacetime obeying the weak energy
condition (that the energy density measured by any timelike
observer is nonnegative).  Since we believe that ``ordinary''
matter obeys the weak energy condition, this theorem demonstrates
that (in the absence of singularities) a time machine cannot be
constructed in a local region --- any attempt to do so will
either fail, or render $B$ non-compact by creating a singularity
(or both).

The theorems of Tipler \rf{5} and Hawking \rf{4} reach slightly
different conclusions from slightly different assumptions.
Tipler assumes that there are tidal forces somewhere on the
Cauchy horizon inside $B$. Specifically, he requires at least one
point $q$ on $H^+(S)\cap B$ at which $K^\mu K^\nu
K_{[\sigma}R_{\rho]\mu\nu [\lambda}K_{\tau ]} \neq 0$, where
$K^\mu$ is the tangent vector to the generator of $H^+(S)$ at
$q$, and square brackets denote antisymmetrization. (The
connection between this condition and tidal forces is discussed
in \Ref{3}.) Hawking, on the other hand, assumes that the
universe is open --- {\it i.e.}, that the surface $S$ is
noncompact --- without making any assumption about the existence
of tidal forces.  Thus, neither theorem applies to Taub-NUT
space, which features a compactly generated Cauchy horizon
without violating the weak energy condition, but which describes
a closed universe free of tidal forces.

Both Tipler's and Hawking's theorems were originally formulated
in the context of (3+1)-dimensional general relativity, but we
may easily extend their analysis to the (2+1)-dimensional case.
The only aspect of the proof that depends on the number of
spacetime dimensions is the use of the (Newman-Penrose) optical
scalar equations.  These equations describe the behavior of a
congruence of null geodesics in terms of scalar quantities,
rather than the tensor quantities that appear in the Jacobi
equation. In (3+1) dimensions four scalars are required (the
expansion, vorticity, and two components of shear), while in
(2+1) dimensions only one (the expansion) is needed.  The
Tipler-Hawking proof uses the equation obeyed by the expansion to
show that the generators of a compactly generated Cauchy horizon
must intersect in the past if the WEC is satisfied.  However, it
is a general property of future Cauchy horizons that the
generators cannot intersect in the past, so the theorem is
proven.  The equation governing the behavior of the expansion in
(2+1) dimensions is obtained from the (3+1)-dimensional
expression by omitting the shear and vorticity terms, as may
easily be checked; it is then straightforward to show that the
dimensionally reduced equation also implies that the generators
of a compactly generated Cauchy horizon must intersect in the
past.\fn{14}{We are grateful to Ted Pyne for discussions on this
point.} Therefore, the theorem applies equally in (2+1) or (3+1)
dimensions.

\goodbreak
\bigskip \bigskip
\centerline{\bigbf APPENDIX B: SOME PROPERTIES OF SU(1,1)}
\nobreak
\medskip

In this Appendix we demonstrate two technical properties
concerning SU(1,1) and the embedding of its parameter space that
is introduced in \sec II-C\.  First, we prove the statement made
in the text concerning the conditions under which a $2 \times 2$
matrix belongs to SU(1,1).  Next we demonstrate the group
invariance of the metric described in the text.

To derive the conditions under which a $2 \times 2$ matrix
belongs to SU(1,1), we begin by parameterizing an arbitrary $2
\times 2$ matrix as
  $$T = \gmatrix{a & b \cr c & d \cr} ,\eqno(B.1)$$
where $a$, $b$, $c$, and $d$ are all complex.  As stated in the
text, SU(1,1) consists of those matrices $T$ satisfying
  $$\det T=+1 \eqno(B.2a)$$
and
  $$T^\dagger \eta T= \eta \ ,\eqno(B.2b)$$
where
  $$\eta = \gmatrix{1&0\cr 0&-1\cr} .\eqno(B.3)$$
{}From \eqs (B.2) one has immediately that
  $$\eqalignno{&a^*a - c^*c = 1 &(B.4a)\cr
    &b^*b - d^*d = -1 &(B.4b)\cr
    &b^*a - d^*c = 0  &(B.4c)\cr
\noalign{\hbox{and}}
    &ad - bc = 1 \ .&(B.5)\cr}$$

If \eq (B.4c) is solved for $c$ and the result is inserted into
\eqpar{B.5}, one finds
  $$a (d^*d - b^*b) = d^* \ .$$
Using \eqpar{B.4b}, this reduces to
  $$a = d^* \ .\eqno(B.6)$$
Combining this result with \eq (B.4c), one has
  $$b^* = c \ .\eqno(B.7)$$
Thus, $T$ can be written as
  $$T = \gmatrix{a & b\cr b^* & a^*\cr} ,\eqno(B.8)$$
where from \eq (B.5) we have
  $$a^*a - b^*b = 1 \ .\eqno(B.9)$$
Comparing with the parameterization
  $$T = \gmatrix{w - i t & y + i x\cr
             y - i x & w + i t\cr }  ,\eqno(B.10)$$
used in the text, one sees that $w$, $t$, $x$, and $y$ are all
real and satisfy
  $$-t^2 + x^2 + y^2 - w^2 = -1 \ . \eqno(B.11)$$

Conversely, it is easily shown that if $w$, $t$, $x$, and $y$ are
all real and satisfy \eq (B.11), then the matrix \eqpar{B.10}
belongs to SU(1,1).

Next, we wish to verify that the metric defined in \sec II-C is
group invariant.  A group transformation on the group parameter
space can be defined by mapping each element of the parameter
space to the element obtained by multiplying on the left by a
fixed element of the group, which we call $\tilde T$.  Thus, the
mapping is defined by
  $$\gmatrix{w' - i t' & y' + i x'\cr y' - i x' & w' + i t'\cr }
     = \tilde T \gmatrix{w - i t & y + i x\cr y - i x & w + i
     t\cr } .\eqno(B.12)$$
Note that the metric of \eq (18) can be written as
  $$\eqalign{ds^2 &= -dt^2 + dx^2 + dy^2 - dw^2 \cr
   &= \det (d T) \ ,\cr}\eqno(B.13)$$
where
  $$dT = \gmatrix{dw - i dt & dy + i dx\cr dy - i dx & dw + i
     dt\cr } . \eqno(B.14)$$
Since $\det \tilde T = 1$, it follows immediately that $\det (d
T') = \det (dT)$, so the metric is invariant.  It is similarly
clear that the metric is invariant under multiplication by a
fixed group element on the right.

(It is not needed in our derivation, but it is interesting to
note that the full invariance group of the metric given by \eq
(18) is SO(2,2), for which the Lie algebra is identical to
SU(1,1) $\times$ SU(1,1).  One of the two SU(1,1) subgroups has
generators that are self-dual (in the 4-dimensional
$w$-$t$-$x$-$y$ space), and the other has generators that are
anti-self-dual.  Transformations of the form described by \eq
(B.12) make up one of the SU(1,1) subgroups, while the other
subgroup corresponds to multiplication on the right.)

\goodbreak
\bigskip \bigskip
\centerline{\bigbf REFERENCES}
\nobreak
\medskip

{\references

\ref{1} R. Geroch and G. T. Horowitz, in {\it General Relativity:
an Einstein Centenary Survey}, eds. S. W. Hawking and W. Israel,
(Cambridge Univ. Press, Cambridge, England, 1979).

\ref{2} K. S. Thorne, in {\it General Relativity and Gravitation
1992}, eds. R. J. Gleiser, C. N. Kozameh, and O. M. Moreschi,
(Institute of Physics Publishing, Bristol, England, 1993).

\ref{3} S. W. Hawking and G. F. R. Ellis, {\it The Large Scale
Structure of Spacetime} (Cambridge Univ. Press, Cambridge,
England, 1973).

\ref{4} S. W. Hawking, {\it Phys. Rev. D} {\bf 46}, 603 (1992).

\ref{5} F. J. Tipler, {\it Phys. Rev. Lett.} {\bf 37}, 879
(1976); {\it Ann. Phys.} {\bf 108}, 1 (1977). See particularly
Theorem 3 of the latter paper.

\ref{6} S. L. Shapiro and S. A. Teukolsky, {\it Phys. Rev. D}
{\bf 45}, 2206 (1992).

\ref{7} A. Ori, {\it Phys. Rev. Lett.} {\bf 71}, 2517 (1993).

\ref{8} J. R. Gott, {\it Phys. Rev. Lett.} {\bf 66}, 1126 (1991).

\ref{9} For a review, see B. Allen and J. Simon, {\it Nature}
{\bf 357}, 19 (1992).

\ref{10} C. Cutler, {\it Phys. Rev. D} {\bf 45}, 487 (1992); see
also A. Ori, {\it Phys. Rev. D} {\bf 44}, R2214 (1991).

\ref{11} L. Marder, {\it Proc. R. Soc.} {\bf A261}, 91 (1961);
A. Staruszkiewicz, {\it Acta. Phys. Polon.} {\bf 24}, 734 (1963);
J. R. Gott and M. Alpert, {\it Gen. Rel. Grav.} {\bf 16}, 243
(1984); S. Giddings, J. Abbott, and K. Kucha\v r, {\it Gen. Rel.
Grav.} {\bf 16}, 751 (1984).

\ref{12} S. Deser, R. Jackiw, and G. 't~Hooft, {\it Ann. Phys.}
{\bf 152}, 220 (1984).

\ref{13} S. Cohn-Vossen, {\it Compositio Math.} {\bf 2}, 69 (1935);
see also A. Huber, {\it Comment. Math. Helv.} {\bf 32}, 13
(1957).

\ref{14} S. M. Carroll, E. Farhi, and A. H. Guth, {\it Phys. Rev.
Lett.} {\bf 68}, 263 (1992); Erratum: {\bf 68}, 3368 (1992).

\ref{15} E. Witten, {\it Nucl. Phys.} {\bf B311}, 46 (1988); S.
Carlip, {\it Nucl. Phys.} {\bf B324}, 106 (1989); S. P. Martin,
{\it Nucl. Phys.} {\bf B327}, 178 (1989); D. Lancaster and N.
Sasakura, {\it Class. Quantum Grav.} {\bf 8}, 1481 (1991).

\ref{16} S. Deser, R. Jackiw, and G. 't~Hooft, {\it Phys. Rev.
Lett.} {\bf 68}, 267 (1992).

\ref{17} G. 't~Hooft, {\it Class. Quantum Grav.} {\bf 9}, 1335
(1992); {\it Class. Quantum Grav.} {\bf 10}, 1023 (1993).

\ref{18} S. Helgason, {\it Differential Geometry and Symmetric
Spaces} (Academic Press, San Diego, 1962); J. A. Wolf, {\it
Spaces of Constant Curvature} (McGraw-Hill, New York, 1967).

\ref{19} J. Balog, L. O'Raifeartaigh, P. Forg\'acs, and A. Wipf,
{\it Nucl. Phys.} {\bf B325}, 225 (1989).

\ref{20} L. Puk\'ansky, {\it Mathematische Annalen} {\bf 156}, 96
(1964).

\ref{21} R. Wald, {\it General Relativity} (University of Chicago
Press, Chicago, 1984).

\ref{22} R.P. Geroch, {\it J. Math. Phys.} {\bf 11}, 437 (1970);
see also \Ref{3}, especially proposition 6.6.8.

\ref{23} C. Nash and S. Sen, {\it Topology and Geometry for
Physicists} (Academic Press, London, 1983).

\ref{24} V. P. Frolov and I. D. Novikov, {\it Phys. Rev. D}
{\bf 42}, 1057 (1990); J. Friedman, M. S. Morris, I. D. Novikov,
F. Echeverria, G. Klinkhammer, K. S. Thorne, and U. Yurtsever,
{\it Phys. Rev. D} {\bf 42}, 1915 (1990); F. Echeverria, G.
Klinkhammer, and K. S. Thorne, {\it Phys. Rev. D} {\bf 44}, 1077
(1991); D. Deutsch, {\it Phys. Rev. D} {\bf 44}, 3197 (1991); J.
L. Friedman and M. S. Morris, {\it Phys. Rev. Lett.} {\bf 66},
401 (1991); J. L. Friedman, N. J. Papastamatiou, and J. Z. Simon,
{\it Phys. Rev. D} {\bf 46}, 4456 (1992); D. S. Goldwirth, M. J.
Perry, T. Piran and K. Thorne, {\it ``The Quantum Propagator for
a Nonrelativistic Particle in the Vicinity of a Time Machine,''}
preprint gr-qc/9308009 (1993).

\ref{25} M. S. Morris, K. S. Thorne, and U. Yurtsever, {\it Phys.
Rev. Lett.} {\bf 61}, 1446 (1988).

\ref{26} S. W. Kim and K. S. Thorne, {\it Phys. Rev. D} {\bf 43},
3929 (1991).

\ref{27} D. G. Boulware, {\it Phys. Rev. D} {\bf 46}, 4421 (1992);
H. D. Politzer, {\it Phys. Rev. D} {\bf 46}, 4470 (1992); M.
Visser, {\it Phys. Rev. D} {\bf 47}, 554 (1993); J. D. E. Grant,
{\it Phys Rev. D} {\bf 47}, 2388 (1993).

\ref{28} H. Waelbroeck, {\it Gen. Rel. Grav.} {\bf 23}, 219 (1991);
{\it Nucl. Phys.} {\bf B364}, 475 (1991).

\ref{29} D. Kabat, {\it Phys. Rev. D} {\bf 46}, 2720 (1992).

\ref{30} P. Menotti and D. Seminara, {\it Phys. Lett.} {\bf B301},
25 (1993); Erratum: {\bf B307}, 404 (1993); P. Menotti and D.
Seminara, {\it Stationary Solutions and Closed Timelike Curves in
2+1 Dimensional Gravity}, preprint hep-th/9305164.

}

\vfill\eject

\ifnum\figurestyle=1

   \nopagenumbers
   \captions
   \centerline{\bf Figure Captions}
   \nobreak
   \medskip

   \capo{1}{\captiona}
   \capo{2}{\captionb}
   \capo{3}{\captionc}
   \capo{4}{\captiond}
   \capo{5}{\captione}
   \capo{6}{\captionf}
   \capo{7}{\captiong}
   \capo{8}{\captionh}
   \capo{9}{\captioni}
\fi

\vfill\eject

\bye